\documentclass[]{agujournal}
\draftfalse

\journalname{JGR-Space Physics}

\usepackage{amsmath}
\usepackage{mathtools}
\usepackage{flafter}
\usepackage{hyperref}

\begin{document}

%
%


\title{Ion velocity and electron temperature inside and around the diamagnetic cavity of comet 67P}

%
%




 \authors{E.~Odelstad\affil{1,2}, A.~I.~Eriksson\affil{1}, F.~L.~Johansson\affil{1}, E.~Vigren\affil{1}, P.~Henri\affil{3}, N.~Gilet\affil{3}, K.~L.~Heritier\affil{4}, X.~Valli\`eres\affil{3}, M.~Rubin\affil{4}, M.~Andr\'e\affil{1}}


\affiliation{1}{Swedish Institute of Space Physics, Uppsala, Sweden.}
\affiliation{2}{Department of Physics and Astronomy, Uppsala University,
Uppsala, Sweden.}
\affiliation{3}{LPC2E, CNRS, Orl\'eans, France}
\affiliation{4}{Department of Physics, Imperial College London, London, United Kingdom}
\affiliation{5}{Physikalisches Institut, Universit{\"a}t Bern, Switzerland.}




\correspondingauthor{E. Odelstad}{elias.odelstad@irfu.se}




\begin{keypoints}
\item The ion velocity exceeded the neutral velocity, showing that the ions were not strongly collisionally coupled to the neutral gas.
\item A population of warm electrons was present throughout the parts of the cavity reached by Rosetta, driving the spacecraft potential negative.
\item A population of cold electrons was consistently observed inside the cavity, and intermittently also in the surrounding region.
\end{keypoints}

This is the pre-peer reviewed version of the following article: \emph{Ion velocity and electron temperature inside and around the diamagnetic cavity of comet 67P} [Odelstad et al., Journal of Geophysical Research: Space Physics, 123, 2018], which has been published in final form at DOI:\href{https://doi.org/10.1029/2018JA025542}{10.1029/2018JA025542}. This article may be used for non-commercial purposes in accordance with Wiley Terms and Conditions for Use of Self-Archived Versions.

%
%


\begin{abstract}
A major point of interest in cometary plasma physics has been the diamagnetic cavity, an unmagnetised region in the inner-most part of the coma. Here, we combine Langmuir and Mutual Impedance Probe measurements to investigate ion velocities and electron temperatures in the diamagnetic cavity of comet 67P, probed by the Rosetta spacecraft. We find ion velocities generally in the range 2-4 km/s, significantly above the expected neutral velocity $\lesssim$1~km/s, showing that the ions are (partially) decoupled from the neutrals, indicating that ion-neutral drag was not responsible for balancing the outside magnetic pressure. Observations of clear wake effects on one of the Langmuir probes showed that the ion flow was close to radial and supersonic, at least w.r.t. the perpendicular temperature, inside the cavity and possibly in the surrounding region as well. We observed spacecraft potentials $\lesssim$-5~V throughout the cavity, showing that a population of warm ($\sim$5~eV) electrons was present throughout the parts of the cavity reached by Rosetta. Also, a population of cold ($\lesssim0.1$~eV) electrons was consistently observed throughout the cavity, but less consistently in the surrounding region, suggesting that while Rosetta never entered a region of collisionally coupled electrons, such a region was possibly not far away during the cavity crossings.
\end{abstract}

%
%

%


%
%
%
%


\section{Introduction}
\label{sec:intro}

\subsection{The Rosetta mission}

Between August 2014 and September 2016, the European Space Agency's Rosetta spacecraft followed the short-period, Jupiter Family comet 67P/Churyumov-Gerasimenko in its orbit around the Sun \citep{Glassmeier2007,Taylor2017}. The comet heliocentric distance ranged from 3.5 au at arrival of the spacecraft, to 1.26 au at perihelion in September 2015, to 3.83 au at the end of mission (EOM). The cometocentric distance of the spacecraft was typically on the order of a few tens to a few hundreds of kilometers, providing unprecedented access to the inner coma of a comet. The relative speed of the spacecraft w.r.t.\ the nucleus was generally on the order of one meter per second or less. Previous cometary space missions (e.g.\ ICE at 21P/Giacobini-Zinner \citep{Rosenvinge1986}, Giotto, Sakigake/Suisei and VEGA (1\&2)  at 1P/Halley \citep{Reinhard1986,Hirao1987,Sagdeev1987} and Giotto at 26P/Grigg-Skjellerup \citep{Grensemann1993}, which all carried plasma instruments) have been short flybys at distances of at least a few hundred km (and relative speeds of tens of kilometers per second). Thus, the Rosetta mission was unprecedented also with regard to its prolonged stay at the target comet, for the first time allowing the long-term evolution of a comet to be observed by in-situ measurements.

\subsection{The diamagnetic cavity}

A major point of interest in cometary plasma physics has been the existence, extent and formation mechanism of the diamagnetic cavity, a region in the inner-most part of the coma into which the interplanetary magnetic field cannot reach and which, in the absence of an intrinsic magnetic field of the nucleus \citep{Auster2015}, will be magnetic-field-free. First predicted theoretically by \citet{Biermann1967}, it has since been observed in situ by the Giotto spacecraft at comet 1P/Halley \citep{Neubauer1986}. \citeauthor{Biermann1967} proposed a pressure balance between the magnetic pressure on the outside of the cavity and the ion dynamic pressure on the inside to account for its formation and extent. However, \citet{Ip1987} and \citet{Cravens1986,Cravens1987} found this to be insufficient to explain the extent of the cavity observed at Halley and instead invoked the ion-neutral drag force inside the cavity to balance the outside magnetic pressure. This was supported by observations of near-equal ion and neutral velocities inside the cavity ($\sim$1~km/s and $\sim$0.9~km/s, respectively), consistent with strong ion-neutral collisional coupling, and clear stagnation of the ion flow in the region just outside the cavity boundary \citep{Balsiger1986,Krankowsky1986}.

A diamagnetic cavity was first detected around comet 67P in magnetometer (RPC-MAG \citep{Glassmeier2007}, hereafter MAG) data from July 26, 2015 \citep{Goetz2016a}, near perihelion at a distance of 170 km from the nucleus (in the terminator plane). The spacecraft remained inside the cavity for about 25 min during this event. Subsequent analysis has identified a total of 665 cavity crossings in MAG data \citep{Goetz2016b}, between April 2015 and February 2016 (i.e.\ some preceding the original detection). They ranged in duration from 8 s up to 40 min, in distance to the nucleus from 40 to 380 km and in heliocentric distance from 1.25-2.4 au. The low velocity of Rosetta ($\lesssim 1$ m/s) implies that these highly transient events were the result of the cavity expanding and contracting over Rosetta's position, rather than resulting from the spacecraft moving into and out of a stationary cavity. Another possibility is blobs of unmagnetized plasma detaching from the main cavity structure and convecting past the spacecraft. The distance to the nucleus of the cavity crossings exhibited a strong statistical dependence on the long-term production rate, but was unaffected by diurnal variations and short-duration events such as outbursts or varying solar wind consitions. \citet{Goetz2016a} therefore suggested a Kelvin-Helmholtz type instability, driven by a presumed velocity shear  at the cavity boundary, to account for its short-term dynamics. This was also proposed to explain the fact that cavity distances were generally found to be larger than predicted for a steady-state cavity sustained by the above pressure balance, as in hybrid simulations by \citet{Koenders2015} and \citet{Rubin2012}. The existence of instabilities at the cavity boundary was indeed confirmed in these simulations.

Density measurements by the Mutual Impedance Probe (RPC-MIP \citep{Trotignon2007}, hereafter MIP) inside the diamagnetic cavity showed densities ranging from $\sim$100 to $\sim$1500 cm$^{-3}$ on longer time scales, but that were almost constant inside any given cavity or between closely successive events \citep{Henri2017}. The surrounding regions of magnetized plasma were in contrast characterized by large density variations, predominantly in the form of large-amplitude compressible structures with relative density fluctuations $\delta n/n \sim$ 1 \citep{Harja2018}. These generally matched similar structures observed in the magnetic field near the cavity by \citet{Goetz2016a}. The plasma density inside the cavity was found to be entirely determined by the ionization of the cometary neutral atmosphere and the cavity boundary generally located close to the electron exobase. Hence, \citeauthor{Henri2017} suggested that the cavity formation and extent was the result of electron-neutral collisionality rather than the ion-neutral collisionality previously invoked. They also proposed a Rayleigh-Taylor type instability of the cavity boundary, driven by the electron-neutral drag force acting as an "effective gravity", instead of the Kelvin-Helmholtz type suggested by \citet{Goetz2016a}. 

\citet{Timar2017} obtained good fits of observed cavity distance values to the ion-neutral drag model of \citet{Cravens1986,Cravens1987}, using the cometary neutral production rate and solar wind dynamic pressure estimated from magnetic field data as well as several different solar wind propagation models. This possibly eliminates the need for an instability at the cavity boundary to account for the intermittent nature of the cavity crossings, in favour of variations in the solar wind pressure.

\citet{Nemeth2016} found that accelerated electrons in the 100~eV range, typically present in the inner coma, were absent inside the diamagnetic cavity, suggesting that these electrons were bound to the field lines and therefore excluded from the cavity.

\subsection{Neutral gas velocity}

For the first few months at the comet, through fall of 2014 up to early 2015, empirical or semi-empirical estimates of the expansion velocity of the neutral coma gas are available from many different sources: doppler shift of the spectral lines of water observed by the Microwave Instrument on the Rosetta Orbiter (MIRO, \citet{Gulkis2007}) \citep{Lee2015,Biver2015,Gulkis2015}, simulation outputs of DSMC models \citep{Bieler2015} constrained by ROSINA-DFMS data \citep{Fougere2016a,Fougere2016b} and direct measurements by ROSINA-COPS ram and nude gauges \citep{Tzou2017}. They all typically give terminal velocities at a few kilometers from the nucleus surface in the range 400~-~800~m/s, with generally a positive correlation between velocity and local outgassing intensity. However, from the period between April 2015 and February 2016 considered in this Paper, published measurements are scarce. \citet{Marshall2017} used a range of 400~m/s to 1~km/s, with a preferred value of 700~m/s, to obtain local effective production rates of H$_2$O from H$_2^{16}$O/H$_2^{18}$O line area ratios obtained by MIRO for the entire period from August 2014 to April 2016, though no more specific values are given from within this period. \citet{Heritier2017} used a one-dimensional model for the neutral gas based on an adiabatic fluid expansion around the nucleus driven by inner boundary conditions on gas outflow velocity from \citet{Huebner2000} and temperature from the thermophysical model of \citet{Davidsson2005} to find terminal velocities of about 800~m/s. For the purpose of comparison to observed ion velocities in this study, we note simply that the neutral outgassing velocity is on the order of 1~km/s, and that exact values are likely to be lower than this estimate rather than higher.

\subsection{Ion velocity}

The primary ionization processes in the cometary coma, photoionization and electron impact ionization \citep{Vigren&Galand2013,Galand2016}, produce ions that are initially cold and flowing with the neutral gas. This is an effect of conservation of momentum: the momentum of the ionizing particle is minuscule compared to that of the much heavier neutral molecule and therefore does not affect its motion in any significant way. The excess energy from the ionization instead goes to the electrons, which are therefore born warm ($T_{\textnormal{e}}$ $\sim$ 10 eV) \citep{Häberli1996,Galand2016}. If there is no electric field, or if the ions are strongly collisionally coupled to the neutrals,  the ions can thus be expected to be cold and flowing with the neutral gas. In the presence of a magnetic field of solar wind origin, the assumption of no electric field fails because of the existence of a convective electric field, which will cause the ions to gyrate, $\mathbf{E} \times \mathbf{B}$ drift and eventually become dynamically part of the solar wind flow (which will be decelerated and deflected by mass-loading) \citep{Coates2004,Szego2000}. This so called \emph{ion pick-up} process takes place over spatial scales on the order of the ion gyro-radius, which for singly charged water group ions ($m_{\textnormal{i}} \approx 18$) in a cometary plasma with typical magnetic field strength $\lesssim$20 nT \citep{Goetz2017} is $\gtrsim$10 km for ion velocities $\gtrsim$1 km/s. Inside the diamagnetic cavity, or at distances outside of it smaller than about 10 km, this process is therefore unimportant for the ion motion. However, the presence of warm electrons (to be further discussed below) suggests the existence of an ambipolar electric field (at least inside the diamagnetic cavity) of a strength on the order of $k_{\textnormal{B}} T_{\textnormal{e}}/q_{\textnormal{e}}r$ to maintain quasi-neutrality of the radially expanding cometary plasma. In such a case, strong collisional coupling to the neutrals is necessary if the ions are to remain at the neutral velocity. Estimates of the location of the ion-neutral collisionopause by \citet{Mandt2016} suggested that Rosetta was generally in a region where ion-neutral collisions were important. However, these estimates did not take into account the reduced collisionality of accelerated ions due to the cross-section for ion-neutral collisions decreasing with energy. \citet{Vigren2017} used a 1D model to simulate the radial acceleration of water group ions interrupted by collisions (primarily charge transfer processes) with neutral water molecules, taking into account the energy-dependence of the cross-sections. They found that for an outgassing rate $\sim$2$\cdot 10^{28}$ s$^{-1}$, typical of 67P near perihelion, even a weak electric field of 0.03 mV/m, typical of what would be expected for an ambipolar field, is sufficient to partially decouple the ions from the neutrals, giving a bulk ion velocity of about 4 km/s at distances $\sim$200 km from the nucleus, typical of the Rosetta spacecraft around perihelion. For an outgassing rate $\sim$2$\cdot 10^{29}$ s$^{-1}$, typical of Halley during the Giotto encounter, collisional coupling was found to prevail. \citet{Vigren2017b} combined  Langmuir probe (RPC-LAP \citep{Eriksson2007}, hereafter LAP) and MIP measurements to produce estimates of the ion velocity for a three-day period near perihelion, including one diamagnetic cavity crossing. They obtained values typically in the range 2-8~km/s, roughly in line with the predictions of \citet{Vigren2017}, lending further support to the supposition that ions are collisionally decoupled from the neutrals at 67P. The presence of an ambipolar electric field, the velocity of the ions and the formation and dynamics of the diamagnetic cavity at 67P are at present poorly understood. In this Paper we attempt to shed some light on these issues by using the method of combined LAP and MIP measurements to produce estimates of the ion velocity throughout the diamagnetic cavity and compare to the surrounding region.

\subsection{Electron temperature}

\citet{Odelstad2015,Odelstad2017} presented measurements of the spacecraft potential ($V_{\textnormal{S/C}}$) by LAP, showing that $V_{\textnormal{S/C}}$ was mostly negative throughout Rosetta's stay at the comet, often below -10 V and sometimes below -20 V. This was attributed to a population of warm ($\sim$5-10 eV) coma photoelectrons, whose presence was explained by the neutral gas not being dense enough to effectively cool these electrons (which are born warm, as mentioned above) by collisions. Positive spacecraft potentials ($\sim$0-5 V) were only observed in regions of very low electron density ($\lesssim$10 cm$^{-3}$), typically far from the nucleus or above the more inactive areas on it, where the positive $V_{\textnormal{S/C}}$ could be explained by low density rather than temperature and where significant electron cooling by neutrals was not possible. Thus, it was concluded that such warm electrons were persistently present in the parts of the coma reached by Rosetta, most notably also around perihelion, where the elevated neutral density would perhaps have been expected to effectively cool the electrons. The statistical nature of this study could not rule out the existence of some brief events of low spacecraft potential hiding in the data set, which would indicate the near-absence of warm electrons. In this paper, we examine this in detail for the diamagnetic cavity crossings and discuss the implications for the physics of the cavity.

In addition to these warm electrons, clear signatures of cold ($\lesssim$0.1 eV) electrons have also been observed by LAP \citep{Eriksson2017} and MIP \citep{Gilet2017}. In LAP, these show up in high-time-resolution current measurements at fixed bias voltage in the form pulses of typical duration between a few seconds and a few minutes, and in bias voltage sweeps in the form of very steep slopes in the current-voltage curve at high positive bias voltages (to be discussed further below). In MIP, they produce a second resonance in the mutual impedance spectra below the total plasma frequency. Since local electron cooling was negligible as evidenced by the presence of warm electrons, this cold plasma was inferred to have formed in a region closer to the nucleus than reached by Rosetta. Together with the intermittent nature of the signatures in LAP data, this was taken as evidence for strong filamentation of the cold plasma close to the nucleus, with individual filaments extending far outside the collisionally dominated region, perhaps even detaching from it entirely. Similar structures were observed to develop in connection to the diamagnetic cavity in global 3D hybrid simulations by \cite{Koenders2015} and have been proposed to result from the instability of the cavity boundary (e.g.\ \citet{Henri2017}). However, observations of cold electrons are not limited to the diamagnetic cavity so the relationship (if any) between the structure and dynamics of cold and unmagnetized plasma, respectively, is not clear. In this paper, we investigate this by examining in detail the presence of cold electrons in the diamagnetic cavity and the surrounding region.

\section{Instrumentation and measurements}
\label{sec:instruments}

The Rosetta spacecraft carried a suite of 5 plasma instruments, the Rosetta Plasma Consortium (RPC) \citep{Carr2007}, including a Langmuir probe instrument (LAP) \citep{Eriksson2007} and a Mutual Impedance Probe (MIP) \citep{Trotignon2007} which are of primary importance in this paper.

\subsection{LAP}
\label{sec:LAP}

\subsubsection{Physical characteristics}

\begin{figure}[t]
	\center
	\includegraphics[width=\textwidth]{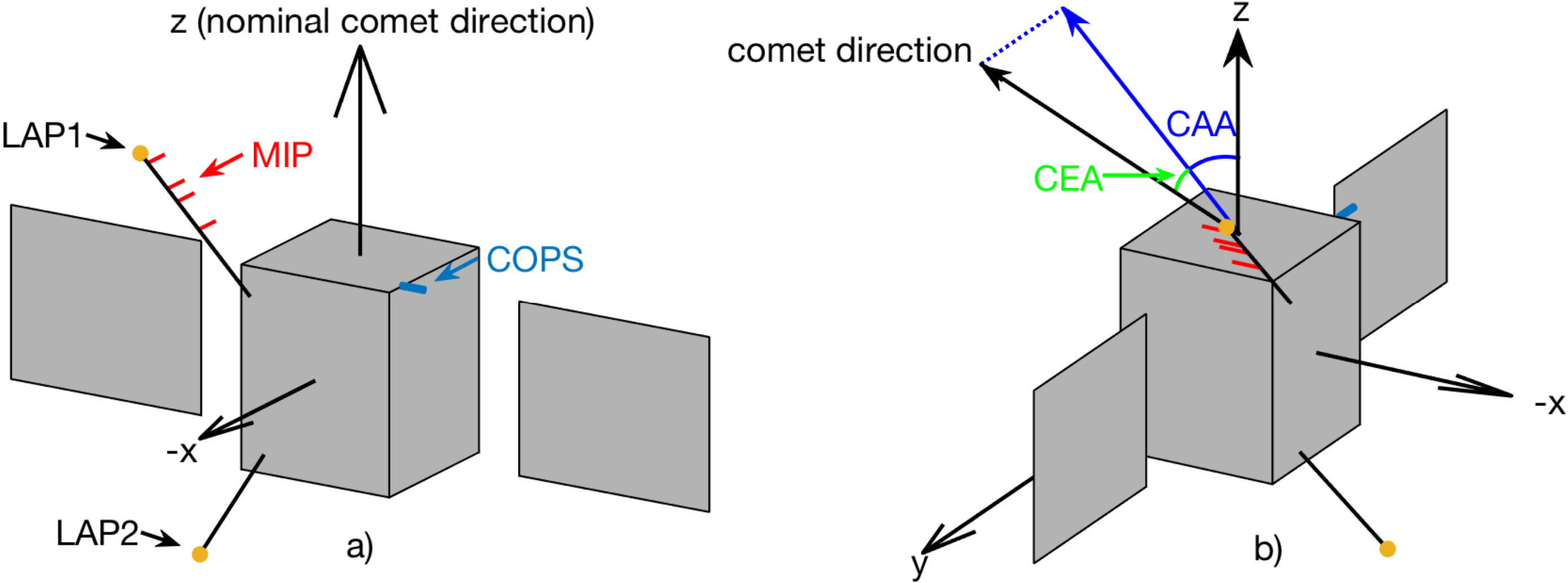}
	\caption{a) Geometrical configuration of the LAP (yellow), MIP (red) and COPS (light blue) sensors. b)~Angles used to describe the spacecraft attitude.}
	\label{fig:SCgeo}
\end{figure}

LAP comprises two spherical Langmuir probes, which we denote LAP1 and LAP2. They are both 5 cm in diameter and have a surface coating of titanium nitride (TiN). Each probe is mounted at the end of a stiff boom protruding from the spacecraft main body, see Figure \ref{fig:SCgeo}a.
 The LAP1 boom is 2.24 m in length and extends from near one of the nominally comet-pointing corners of the shadow ($-x$) side of the spacecraft (the side on which the lander was originally mounted) at an angle of 45$^\circ$ from the nominal comet-pointing direction. The LAP2 boom (also known as the MAG boom, since it also hosts the magnetometer sensors) is 1.62 m in length and extends from near the corner "downstream" of LAP1 on the shadow side, at an angle of about 120$^\circ$ from the nominal comet-pointing direction. Thus, LAP2 is likely to be much more susceptible to wake effects than LAP1.
 
For future reference, Figure \ref{fig:SCgeo}b introduces the Comet Elevation Angle (CEA) and Comet Aspect Angle (CAA) to describe the spacecraft attitude w.r.t.\ the comet nucleus. These are the elevation and azimuth angles, respectively, of the comet position vector in a spherical coordinate system with zenith direction along the spacecraft y-axis and azimuth reference direction along the spacecraft x-axis. Thus, CAA is the angle of the projection of the nucleus position vector in the x-z plane from the z-axis (in the range [-180$^\circ$,180$^\circ$], positive for positive $x$) and CEA is the angle between the comet position vector and this projection (in the range [-90$^\circ$,90$^\circ$], positive for positive $y$). The corresponding angles for the Sun are denoted SEA and SAA, respectively.

\subsubsection{Operational modes}

LAP supports three main operating modes: current measurements at fixed bias potential, potential measurements at fixed bias current (or with a floating probe, i.e.\ disconnected from the biasing circuitry) and Langmuir probe bias potential sweeps. In the latter mode, which is the one used in this paper, the bias voltage is sequentially stepped through a range of values and the current sampled at each step. The resulting $I$-$V$ curve can the be used to derive various parameters of the surrounding plasma by comparison to theoretical models for the probe current-voltage relationship. LAP supports bias potentials between -30 and +30 V, with typical voltage steps between 0.25 and 1 V. Full sweeps typically take between 1 and 4 s and are performed at a cadence of 64-160 s.

\subsubsection{Theoretical models}

The theory of current collection by spherical (and cylindrical) Langmuir probes immersed in a plasma was pioneered by \citet{Mott-Smith1926} for the case of a stationary maxwellian plasma in the \emph{orbit-motion limited} (OML) regime, where the Debye length $\lambda_\textrm{D}$ is much smaller than the radius of the probe. For an ideal isolated spherical probe, the expression reads

\begin{linenomath*}
\begin{equation}
	I_{j} = \left\{ \begin{array}{ll}
		I_{j\textnormal{0}} ( 1 - \chi_j), & \chi_j \leq 0, \\
		I_{j\textnormal{0}} \exp \left\{-\chi_j \right\}, & \chi_j \geq 0. \end{array} \right.
\label{eq:OML_I}
\end{equation}
\end{linenomath*}
where $j$ denotes the particle species (i for ions and e for electrons, respectively) and the random thermal current $I_{j\textnormal{0}}$ and normalized potential $\chi_j$ are given by

\begin{linenomath*}
\begin{eqnarray}
	I_{j\textnormal{0}} & =  & -4\pi a^2 n_j q_j \sqrt{\frac{k_{\textnormal{B}} T_j}{2\pi m_j}}, \label{eq:OML_I0} \\
	\chi_j & = & \frac{q_j V_{\textnormal{p}}}{k_{\textnormal{B}} T_j}. \label{eq:OML_Xi}
\end{eqnarray}
\end{linenomath*}
Here, $a$ is the probe radius, $V_{\textnormal{p}}$ the probe potential w.r.t.\ the ambient plasma and $n_j$, $q_j$, $T_j$ and $m_j$ are, respectively, the number density, charge, temperature (in Kelvin) and mass of particle species $j$. LAP uses the spacecraft as electrical ground, thus $V_{\textnormal{p}}$ is related to the controlled bias potential $U_{\textnormal{B}}$ as $V_{\textnormal{p}} = U_{\textnormal{B}} + V_{\textnormal{S/C}}$, where $V_{\textnormal{S/C}}$ is the spacecraft potential. In the case of Langmuir probe bias potential sweeps where $U_{\textnormal{B}}$, and thus $V_{\textnormal{p}}$, is incrementally varied on timescales short enough that all other parameters can be assumed to be constant, the probe current due to species $j$ will be linearly proportional to $V_{\textnormal{p}}$, and thus $U_{\textnormal{B}}$, for attractive bias potentials $\chi_j \leq 0$ (i.e.\ electrons and a positively charged probe or positive ions and a negative probe). The proportionality constant, hereafter referred to as the ion or electron slope, is given by

\begin{linenomath*}
\begin{equation}
	\frac{\partial I_j}{\partial U_{\textnormal{B}}} = \frac{\partial I_j}{\partial V_{\textnormal{p}}} =  a^2 n_j q_j^2 \sqrt{\frac{8\pi}{m_j k_{\textnormal{B}} T_j}}.
\label{eq:eslope}
\end{equation}
\end{linenomath*}
Equation (\ref{eq:eslope}) for the electron slope is used in this paper together with empirical electron slopes from LAP sweeps and density measurements from MIP to constrain the electron temperature and investigate the respective prevalences of warm and cold electrons in the diamagnetic cavity and the surrounding region (cf.\ Section \ref{sec:res}).

For repulsive bias potentials $\chi_j \geq 0$ (i.e.\ electrons and a negatively charged probe or positive ions and a positive probe), the probe current falls off exponentially with increasing $|V_{\textnormal{p}}|$. Due to the larger ion mass, the random thermal ion current $I_{\textnormal{i0}}$ is much smaller than the random thermal electron current $I_{\textnormal{e0}}$, thus the ion current is typically negligible at repulsive bias potentials. The electron current on the other hand is generally not negligible unless $-V_{\textnormal{p}} \gtrsim k_{\textnormal{B}} T_{\textnormal{e}}/q_{\textnormal{e}}$. Probe currents at repulsive bias potentials are not utilized in this paper other than to note that ion and electron slopes have to be determined from sweep regions at sufficiently large $|V_{\textnormal{p}}|$ that the current due to the oppositely charged species is effectively suppressed.

In addition to currents due to collection of ambient plasma particles, particle emission from the probe surface in the form of photoelectrons and secondary electrons in response to solar EUV and impacting plasma electrons, respectively, may also contribute to the probe current. Attraction or repulsion of photoelectrons is determined by the probe potential w.r.t.\ its immediate surroundings, which may differ from $V_{\textnormal{p}}$ if part of the spacecraft potential field persists at the position of the probe. For a sunlit probe the photoelectron current will be independent of the bias potential at $U_{\textnormal{B}} < -\alpha V_{\textnormal{S/C}}$, where $\alpha$ is the fraction of $V_{\textnormal{S/C}}$ remaining at the probe position, since all emitted electrons are repelled and escape from the probe, contributing to the current. At $U_{\textnormal{B}} > -\alpha V_{\textnormal{S/C}}$, the photoelectron current falls off exponentially with increasing $U_{\textnormal{B}}$, with a characteristic e-folding typically on the order of 1-2 V. This regime change at $U_{\textnormal{B}} = -\alpha V_{\textnormal{S/C}}$ is typically identifiable as a sharp knee in the sweeps, hereafter denoted $V_{\textnormal{ph}}$, from which an estimate of the spacecraft potential can be obtained as $V_{\textnormal{S/C}} = -V_{\textnormal{ph}}/\alpha$. $\alpha$ has been shown to generally be in the range 0.7-1 by \citet{Odelstad2017}. Such spacecraft potential measurements have previously been used by \citet{Odelstad2015,Odelstad2017} to demonstrate the overall pervasiveness of warm ($\sim$5 eV) electrons in the coma of 67P; in this paper we more carefully examine this during the times when the spacecraft is inside the diamagnetic cavity.

The OML theory for spherical probes was extended to the case of a drifting maxwellian plasma by \citet{Medicus1961}. An analytically simpler semi-empirical approximation to the rather cumbersome expressions of \citeauthor{Medicus1961} was presented by \citet{Fahleson1967} for attractive probe potentials (e.g.\ positive ions and a negatively charged probe). Here, the attracted-ion current is still given by Equation (\ref{eq:OML_I}) but with modified expressions for $I_{\textnormal{i0}}$ and $\chi_{\textnormal{i}}$:

\begin{linenomath*}
\begin{eqnarray}
	I_{\textnormal{i0}} & = & -4\pi a^2 n_{\textnormal{i}} q_{\textnormal{i}} \sqrt{\frac{k_{\textnormal{B}} T_{\textnormal{i}}}{2\pi m_{\textnormal{i}}} + \frac{v_{\textnormal{D}}^2}{16}} \label{eq:FahlesonI0}\\
	\chi_{\textnormal{i}} & = & q_{\textnormal{i}} U_{\textnormal{B}} \Big/ \left( k_{\textnormal{B}} T_{\textnormal{i}} + \frac{m_{\textnormal{i}} v_{\textnormal{D}}^2}{2} \right), \label{eq:FahlesonX}
\end{eqnarray}
\end{linenomath*}
where $v_{\textnormal{D}}$ is the drift velocity of the ions. This gives for the ion slope:

\begin{linenomath*}
\begin{equation}
	\frac{\partial I_{\textnormal{i}}}{\partial U_{\textnormal{B}}} = \frac{2\pi a^2 n_{\textnormal{i}} q_{\textnormal{i}}^2}{m_{\textnormal{i}}} \underbrace{\frac{\sqrt{\frac{8k_{\textnormal{B}} T_{\textnormal{i}}}{\pi m_{\textnormal{i}}} + v_{\textnormal{D}}^2}}{\frac{2k_{\textnormal{B}} T_{\textnormal{i}}}{m_{\textnormal{i}}} + v_{\textnormal{D}}^2}}_{1/v_{\textnormal{i}}},
\label{eq:ionslope}
\end{equation}
\end{linenomath*}
where for convenience, the factor containing the dependence on ion motion ($T_{\textnormal{i}}$ and $v_{\textnormal{D}}$) has been denoted $v_{\textnormal{i}}$. It should be noted at this point that in the case of a stationary plasma ($v_{\textnormal{D}} \rightarrow 0$) Equations (\ref{eq:FahlesonI0}) - (\ref{eq:FahlesonX}) reduce to the expressions of Equations (\ref{eq:OML_I0}) - (\ref{eq:OML_Xi}) and $v_{\textnormal{i}}$ reduces to $\frac{\sqrt{\pi}}{2} v_{\textnormal{th}} \approx v_{\textnormal{th}}$, where $v_{\textnormal{th}} = \sqrt{2 k_{\textnormal{B}} T_{\textnormal{i}} / m_{\textnormal{i}}}$ is the thermal velocity of the ions, defined as their most probable speed. In the case of cold drifting ions ($T_{\textnormal{i}} \rightarrow 0$), $v_{\textnormal{i}}$ reduces to the drift velocity $v_{\textnormal{D}}$. Thus, $v_{\textnormal{i}}$ represents an effective ion velocity that combines the effects of thermal and drift motions of the ions on the probe current collection, the contributions of which cannot be separated by Langmuir probe measurements alone.

\subsubsection{Practical aspects}

While the theory of spherical probes in a (possibly drifting) single-component maxwellian plasma is well developed, the behavior of such probes in the highly variable and dynamic multi-component (at least in terms of the electron temperature \citep{Eriksson2017}) cometary plasma and in close proximity to a large, negatively charged spacecraft is not well understood. Space-charge sheath and wake effects are to be expected, complicating the analysis and interpretation of probe measurements \citep{Sjogren2012,Johansson2016sctc}. Among the more robust parameters that can reliably be determined from LAP measurements alone are the spacecraft potential \citep{Odelstad2015,Odelstad2017} and the photosaturation current \citep{Johansson2017}. In addition to these, the aforementioned ion slope can also most often be reliably identified in LAP sweeps. While this curve parameter alone cannot determine any parameter of the plasma, if the ion density is known from some other source, Equation (\ref{eq:ionslope}) can be used to obtain the effective ion velocity $v_{\textnormal{i}}$ if assumptions are made on the values of $m_{\textnormal{i}}$ and $q_{\textnormal{i}}$. The cometary plasma is most often dominated by singly charged H$_2$O$^+$ and H$_3$O$^+$ ions \citep{Vigren&Galand2013,Heritier2017}, giving $m_{\textnormal{i}} \approx 18$~u and $q_{\textnormal{i}} = 1$ e. The total plasma density can often be reliably obtained from MIP (cf.\ Section \ref{sec:MIP}). Thus, combining MIP density measurements with the ion slope from LAP sweeps provides a means of measuring the ion velocity in the cometary plasma. This method was used by \citet{Vigren2017b} to obtain ion velocities for a three-day period near perihelion. It must be noted here that this method presumes that the LAP ion slope is unaffected by any sheath or wake effects of the spacecraft. The validity of this assumption, and possible consequences on obtained results when it fails, will be discussed in detail in Section \ref{sec:disc}.

A more detailed discussion of the general appearance and interpretation of LAP1 sweeps can be found in Appendix A, along with descriptions of how the various sweep parameters (e.g.\ ion and electron slopes) are obtained from the sweeps.

\subsection{MIP}
\label{sec:MIP}

The MIP antenna consists of four cylindrical electrodes (1 cm in diameter) arranged in a linear array along the LAP1 boom, see Figure \ref{fig:SCgeo}. The two middle electrodes, 20 cm apart, are transmitting monopoles while the two outer ones, each at a 40 cm distance from the nearest transmitter, make up a receiving electric dipole with a total separation distance of 1 m. In the most commonly used active mode, an sinusoidal current is fed to the transmitting electrodes, either in phase or anti-phase, and the mutual impedance between the transmitting and receiving electrodes is computed from the induced voltage difference across the receiving dipole. This is repeated for a number of different frequencies of the driving sinusoidal current, producing a mutual impedance spectrum from which properties of the surrounding plasma can be deduced. In particular, the modulus of the mutual impedance spectrum exhibits either a peak or a cut-off (depending on the plasma properties) at the electron plasma frequency $f_{\textnormal{p}} = \frac{1}{2\pi}\sqrt{n_{\textnormal{e}} q_{\textnormal{e}}^2 / \varepsilon_{\textnormal{0}} m_{\textnormal{e}}} \approx 9 \sqrt{n_{\textnormal{e}}}$ , where $n_{\textnormal{e}}$ in cm$^{-3}$ gives $f_{\textnormal{p}}$ in kHz, from which the total electron density can thus be obtained. This signature at the plasma frequency appears in the mutual impedance spectra only when the Debye length is small enough compared to the emitter-receiver distance. For longer-Debye-length plasmas, typically when it exceeds about half a meter, LAP2 was used to transmit the MIP oscillating signal into the plasma, still received on the MIP receivers, in order to provide a larger transmitter-receiver distance, close to 5 m, and extend the usable range of the mutual impedance technique to plasmas characterized by a Debye length up to a few meters. This operational mode is called the Long Debye Length (hereafter LDL) mode, while the previously described mode is referred to as the Short Debye Length (hereafter SDL) mode. 

While the theory of mutual impedance probes in a homogeneous maxwellian plasma is well developed, the behavior of such probes in close proximity to a large, negatively charged spacecraft is not well understood. In particular, questions may be raised regarding what density is actually measured by MIP from its location inside the plasma sheath created by the potential field of the charged spacecraft. The local electron density at the position of the probes is expected to be much suppressed by this potential field, especially since the boom on which the electrodes are mounted is conductive and grounded to the spacecraft. In passive mode the picture is clearer: conservation of energy requires the frequency of externally generated waves to be constant during propagation through the density gradient surrounding the spacecraft. The spectral peak has been observed not to change much when switching between active and passive modes, when electrostatic waves not generated by the MIP experiments but by other plasma processes are observed at the plasma frequency (not shown here). Thus, we surmise that also active mode measurements are reliable for obtaining the density of the ambient plasma, unperturbed by the presence of the spacecraft. In fact, all MIP density measurements used in this paper have been obtained from active modes, since MIP has a better power resolution in active than in passive mode. The power resolution in passive mode is generally too low to observe variations of $f_{\textnormal{p}}$, except on some occasions when it confirms the active signature.

\citet{Gilet2017} showed that in the presence of two distinct electron populations at different temperatures, the resonance peak at the total electron plasma frequency observed in the modulus of the mutual impedance spectrum is supplemented by a second resonance peak at lower frequency, close to the plasma frequency of the cold population, corresponding to electron acoustic waves excited in the plasma by the MIP experiment. This allows MIP to detect the presence of such a cold electron population in the cometary plasma, as has indeed been the case for parts of the mission \citep{Gilet2017}. In such cases, the position of the resonance at the total plasma frequency gives the total electron density; this is the density estimate used throughout this paper. In principle, MIP could be used also to investigate the prevalence of cold electrons. However, the technique for doing this systematically with MIP is still under development and in this paper we primarily use LAP for that purpose.

\subsection{Combining LAP and MIP data}
\label{sec:interpolation}

Both MIP and LAP working modes are organized in synchronized 32-s long sequences composed of elementary working modes. The 64-160-s cadence of LAP sweeps implies that sweeps are performed every other to every fifth such 32-s sequence, always at the very beginning of the sequence. MIP begins each 32-s sequence with a full active-mode spectrum, which is then repeated with a cadence of typically just over 4 s, when MIP is in burst mode. Thus, each LAP sweep is concurrent with a MIP spectrum, which in the most common case of a $\sim$3-s sweep from negative to positive bias voltage is generally acquired during the first half of the sweep, i.e.\ simultaneously with the ion side of the sweep. Thus the timing between MIP density measurements and LAP ion slope measurements is generally very good. However, not all spectra are well-behaved enough to allow automated identification of the plasma frequency signature so some sweeps will not have a concurrent density estimate from MIP. In the quiescent plasma inside the diamagnetic cavity, such sweeps are instead combined with density estimates from adjacent spectra, provided that the mid-points are less than 4 s apart. This is justified by the fact that the plasma density has been observed to be almost constant during the diamagnetic cavity crossings \citep{Henri2017}. In the more variable plasma outside the cavity, such sweeps are omitted when producing ion velocity estimates by combining LAP1 ion slopes with MIP densities.

\section{Results}
\label{sec:res}

\subsection{Case study of the Nov 19-21 2015 cluster of diamagnetic cavity observations}
\label{sec:res1}

In order to facilitate comparative statistics between the cavity and the surrounding region, we seek an interval with a comparable number of measurements inside and outside the cavity. It should be short enough for secularly varying parameters such as comet outgassing, latitude and radial distance of the spacecraft to be constant but still long enough to contain sufficient measurements for good statistics. The diamagnetic cavity events came in the form of single events and clusters; the most prominent clusters occurred on 30 July 2015 and 19-21 November 2015 \citep{Goetz2016b}. We found that the 50 h interval between 08:00 on 19 November 2015 and 10:00 on 21 November 2015 fulfilled the above criteria. During this interval the spacecraft spent about one third of the time inside the cavity and both MIP density measurements and LAP sweeps from both probes are available with sufficient quality to perform our study. Latitude and radial distance varied from -40$^\circ$ to -57$^\circ$ and 126~km to 150~km, respectively.

\begin{figure}[t]
	\hspace{-0.2\textwidth}
	\includegraphics[width=1.4\textwidth]{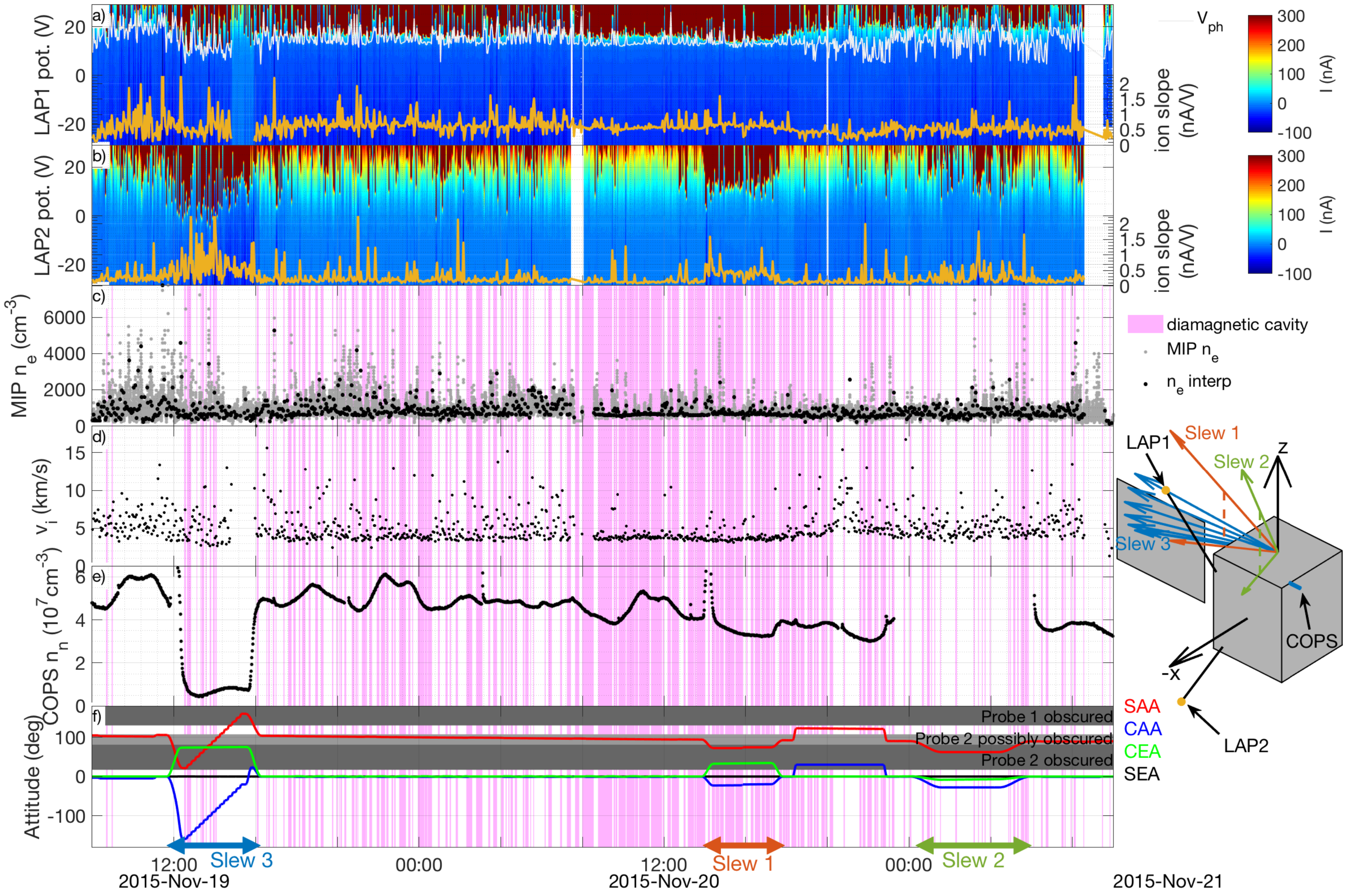}
	\caption{Multi-instrument time series plot of the interval from 08:00 19 November 2015 to 10:00 21 November 2015. See text for description.}
	\label{fig:timeseries}
\end{figure}

Figure \ref{fig:timeseries} shows a multi-instrument time series plot of this interval. 
Figures \ref{fig:timeseries}a and \ref{fig:timeseries}b show sweeps from LAP1 and LAP2, respectively, with probe bias potential on the (left-hand) y-axis and measured probe current color-coded on the surface plot. The color scale is set quite narrow in order to bring out at least the most prominent features on the ion (negative-voltage) side of the sweeps. As a consequence, the much larger electron currents on the positive-voltage side often saturate the color scale. For LAP1, the photoelectron knee, roughly corresponding to the negative of the spacecraft potential, is over-plotted as a white line, to be read off the same y-axis. To more clearly make out the behavior of the sweep ion current, the ion slope calculated from each sweep is plotted as an orange line on top of the negative-voltage side of the sweeps, to be read off the respective right-hand y-axes. Figure \ref{fig:timeseries}c contains scatter plots of all the MIP plasma density measurements acquired during the interval (grey points) and those for which the timing coincide with the timing of LAP sweeps (black points). Here, and in all the remaining panels, time intervals when \citet{Goetz2016b} have identified the spacecraft to be inside the diamagnetic cavity are shaded purple. Figure \ref{fig:timeseries}d shows the ion velocities derived from combining the LAP ion slopes with the simultaneous MIP densities through Equation (\ref{eq:ionslope}). Figure \ref{fig:timeseries}e shows the neutral density measurements by COPS during the plotted interval. Finally, Figure \ref{fig:timeseries}f displays the spacecraft attitude through the angles defined in Figure \ref{fig:SCgeo}b, with dark shaded regions denoting the angular intervals where each of the probes is obscured from view of the sun (SAA) or comet (CAA, CEA) and a lighter shaded region where Probe 2 is possibly shaded behind the High Gain Antenna (HGA), depending on the HGA orientation which we do not look further into here.

\begin{figure}[t]
	\includegraphics[width=\textwidth]{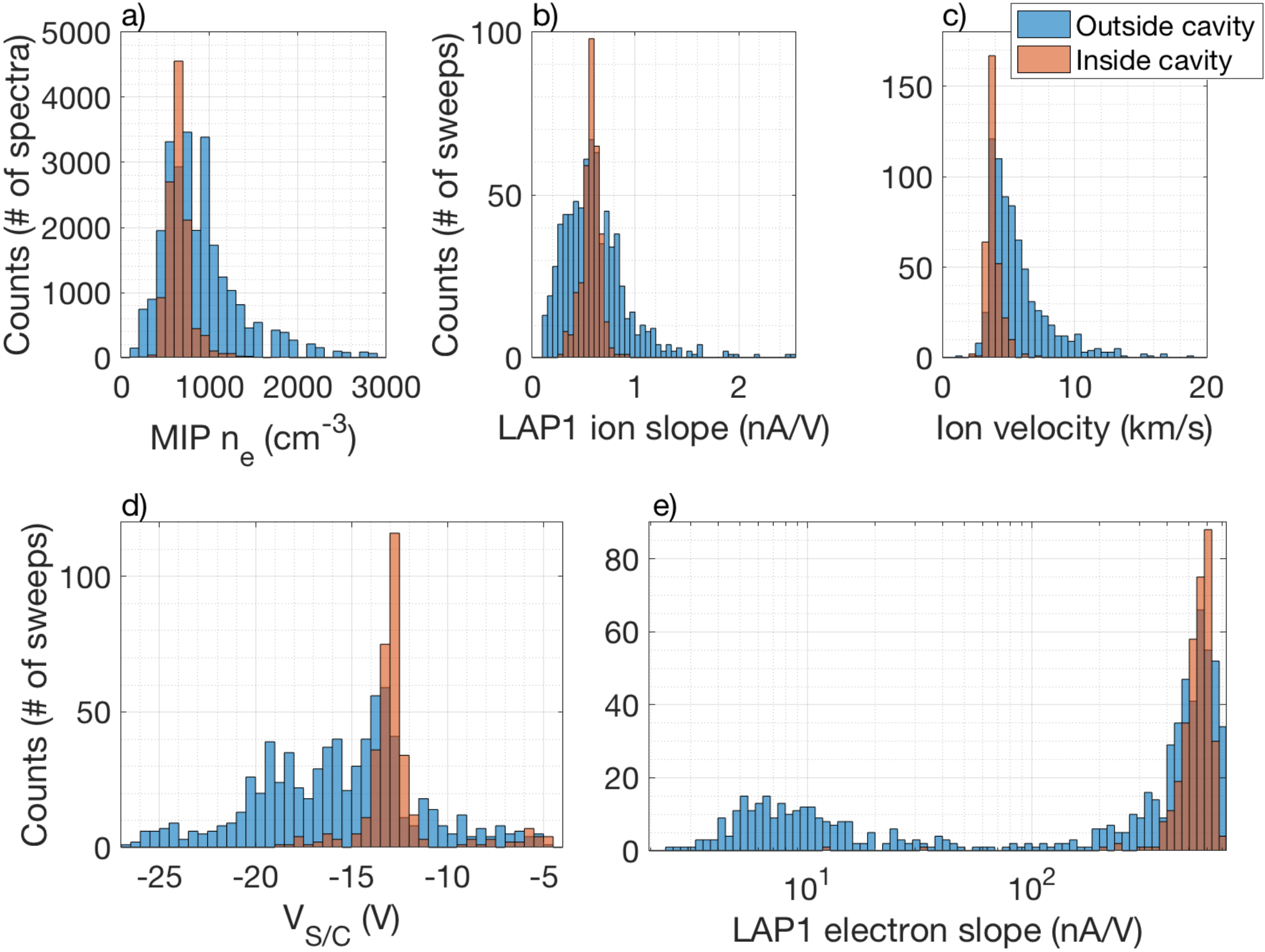}
	\caption{Histograms of data in the interval from 08:00 19 November 2015 to 10:00 21 November 2015.}
	\label{fig:hist}
\end{figure}

Turning first our attention to Figure \ref{fig:timeseries}c, we recall the observation of \citet{Henri2017} that the density is remarkably stable inside the cavity. This is true both during each individual cavity event, but also when comparing densities from different events in this interval. Outside the cavity on the other hand, the density is highly variable, with scatter primarily towards higher densities. This is shown more clearly in Figure \ref{fig:hist}a,
where we show a histogram of the density measurements in the plotted time interval. Here, we also see that the density is quite narrowly and almost symmetrically distributed around 600-700 cm$^{-3}$ inside the cavity, with most measurements falling in the range 500-800 cm$^{-3}$. The measurements from outside the cavity exhibit much larger spread and distribute less symmetrically.

The LAP1 ion slopes in Figure \ref{fig:timeseries}a exhibit similar behavior, as shown more clearly in the histogram in Figure \ref{fig:hist}b. Here too the spread is much larger in the region surrounding the cavity than inside of it. The distribution is less asymmetric outside the cavity than it was for the densities, though some preponderance of lower values may be discerned.

The velocities in Figure \ref{fig:timeseries}d, a histogram of which is shown in Figure \ref{fig:hist}c, display an even larger difference between the cavity and the surrounding region. Inside the cavity, the ion velocities are very narrowly distributed around 3.5-4 km/s, with almost all measurements falling in the range 3-4.5 km/s. In fact, the spread here is so small that it may just be an effect of measurement noise, there being no statistically significant variation of the derived ion velocity at all inside the cavity during this interval. We note specifically that the ion velocity is significantly larger than the neutral velocity of $\lesssim$1 km/s at all times, contrary to what would be expected if the ion-neutral drag force \citep{Cravens1986,Cravens1987} was responsible for balancing the outside magnetic pressure at the cavity boundary. Also, we note from Figure \ref{fig:timeseries}e that the neutral density varies by $\sim$50\% during this time interval (excluding the spacecraft slew between 12:00 and 16:00 on November 19, to be discussed below); this does not come through at all in the ion velocity measurements inside the cavity. The velocity outside the cavity is on the contrary highly variable, with a spread entirely towards higher velocities, up to and above 10 km/s. Here too there is no sign of correlation with the neutral density. We see no sign of a decrease of the ion velocities at the cavity boundary or in the surrounding region. However, our ion velocity estimate cannot distinguish between bulk drift and thermal speeds or different flow directions, so this does not preclude a decrease in the outward radial bulk flow outside the cavity.

A histogram of the spacecraft potential inside and outside the cavity is shown in Figure \ref{fig:hist}d. $V_{\textnormal{S/C}}$ is consistently $< -5$ V both inside and outside the cavity, attesting to the persisting presence of warm ($\sim$5 eV) electrons in both these regions. The measurements distribute narrowly around -13 V inside the cavity, with most measurements falling in the range between -14 and -12 V. Outside the cavity the distribution is much wider, with significant spread primarily towards more negative potentials. In this sense, the spacecraft potential measurements mirror the density measurements in Figure \ref{fig:hist}a, consistent with $V_{\textnormal{S/C}}$ being governed mainly by the electron density.

Figure \ref{fig:hist}e shows a histogram of the LAP1 electron slopes inside and outside the cavity. Outside the cavity, the measurements distribute into two distinct groups, $\lesssim$20 nA/V and $\gtrsim$300 nA/V, respectively. Inside the cavity, only the population $\gtrsim$300 nA/V is observed. The electron slope can be related to the electron temperature and density through Equation (\ref{eq:eslope}). Solving Equation (\ref{eq:eslope}) for $T_{\textnormal{e}}$ in units of eV gives

\begin{equation}
	T_{\textnormal{e}} \textnormal{ [eV]} = 0.04 \cdot n_{\textnormal{e}} \textnormal{ [cm$^{-3}$]} \Bigg/ \frac{\partial I_{\textnormal{e}}}{\partial U_{\textnormal{B}}} \textnormal{ [nA/V]},
\label{eq:eslope2}
\end{equation}
where $n_{\textnormal{e}}$ [cm$^{-3}$] and $\partial I_{\textnormal{e}} / \partial U_{\textnormal{B}}$ [nA/V] are the density in units of cm$^{-3}$ and electron slope in nA/V, respectively. For $n_{\textnormal{e}} \lesssim 800$ cm$^{-3}$ and slopes $\gtrsim$300 nA/V, typical inside the diamagnetic cavity, this gives $T_{\textnormal{e}} \lesssim 0.3$ eV. Thus, such cold electrons are pervasive throughout the cavity. For slopes $\lesssim$20 nA/V, $T_{\textnormal{e}}$ comes out $\gtrsim$2.5 eV for $n_{\textnormal{e}} \gtrsim 150$ cm$^{-3}$, which is the very lowest density ever observed throughout this time interval. This corresponds well to the temperature range inferred for the warm electron population from the spacecraft potential measurements above. Thus, these slopes correspond to sweeps where the cold electron population is not observed and the warm population prevails. For larger densities on the order of several hundreds or thousands of cm$^{-3}$, more typical of what is generally observed in the region surrounding the cavity, Equation (\ref{eq:eslope2}) gives electron temperatures on the order of a few tens to several hundreds of eV, entirely inconsistent with the spacecraft potential measurements. The electron slopes, especially the low ones corresponding to warm-only electrons, are most likely heavily influenced by spacecraft sheath effects. Thus, the temperature estimates obtained from them, especially for the warm ones, should not be taken as a quantitative measure of  $T_{\textnormal{e}}$, but rather as an indication of the presence or absence of the cold electron population in the LAP sweeps.

Several spacecraft slews were performed during the plotted interval, as can be seen in the attitude angles in Figure \ref{fig:timeseries}f (Slew 1, Slew 2 and Slew 3). These are also graphically illustrated in the inset next to Figure \ref{fig:timeseries}d. We focus here on Slew 1, between about 14:00 and 17:40 on November 20, during which the spacecraft was inside the cavity most of the time. Looking at panel b in Figure \ref{fig:timeseries}, we see a clear regime change in the current collection of LAP2 coinciding with this slew. The LAP2 ion slope, generally on the order of 0.2 nA/V during nominal pointing conditions, effectively doubles to $\sim$0.4 nA/V during the slew, bringing it much closer to typical LAP1 values $\gtrsim$0.5 nA/V inside the cavity. This regime change also comes through clearly on the positive-voltage side of the LAP2 sweeps, with a substantial increase of the electron current at bias potentials $\gtrsim$10 V due to collection of cold electrons. Cold electrons are only very rarely seen in LAP2 sweeps before and after this slew, but are clearly and consistently present during the slew, at least inside the cavity. These observations strongly point to LAP2 being in or near a wake created by the spacecraft in the ion flow during nominal pointing, from which it comes out, at least partially, during the slew as the pointing changes and the probe becomes better exposed to the flowing plasma. This explains the lower ion slopes in LAP2 during nominal pointing (compared to both LAP1 and LAP2 during the slew) since the ion flow into a wake would be weakened. It also explains the increase of the cold electron current, since in the presence of warm electrons such a wake in the ion flow would become negatively charged, thereby prohibiting cold electrons from entering. In order for a wake to form, the ion flow must be supersonic, at least w.r.t.\ the perpendicular ion temperature. The observations during this slew thus constitute clear evidence of a directed ion flow inside the cavity. They are consistent with a flow direction radially outward from the nucleus, at least inside the cavity.

\begin{figure}[t]
	\includegraphics[width=\textwidth]{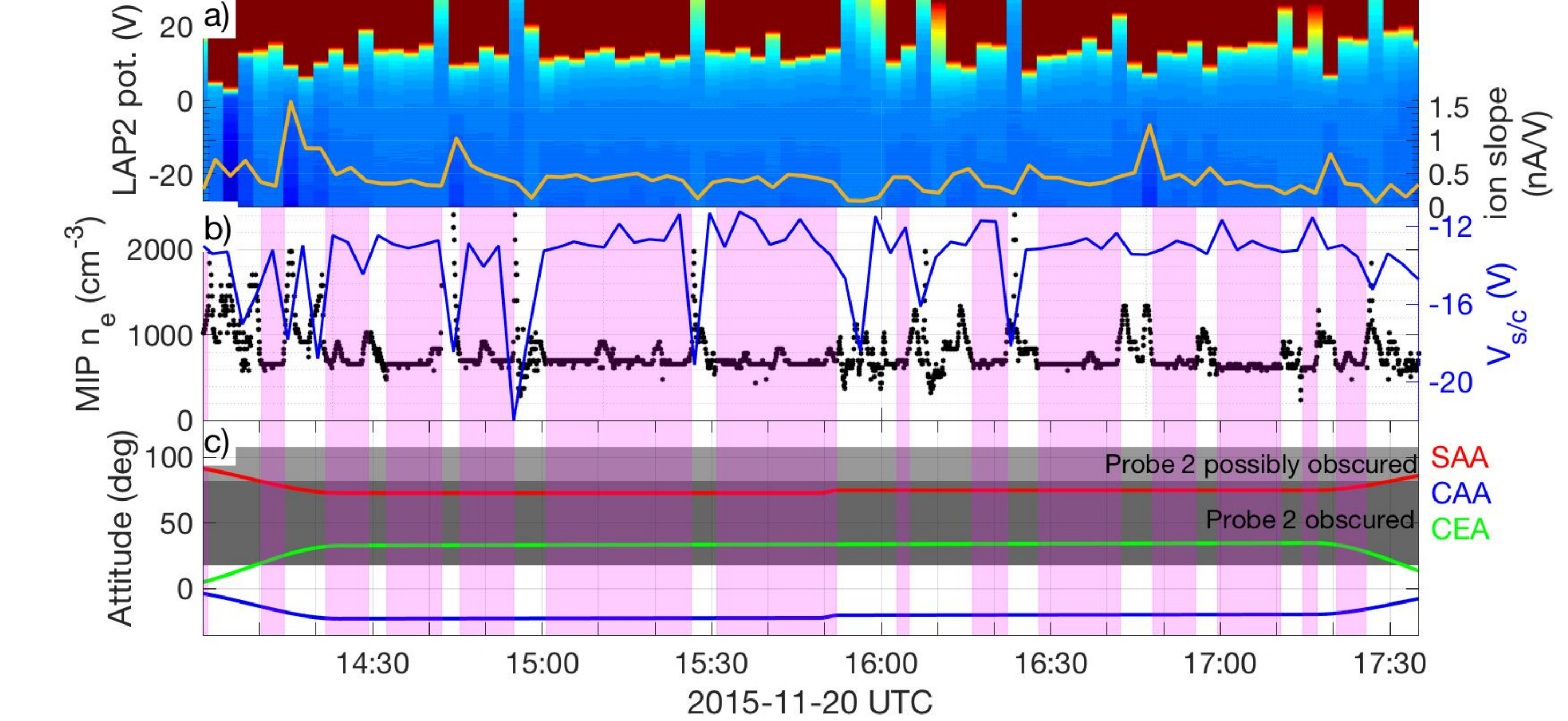}
	\caption{Zoomed-in plot of Slew 1 in Figure \ref{fig:timeseries}. Panels a, b and c are zoom-ins of panels b, c  and f in Figure \ref{fig:timeseries}, respectively.}
	\label{fig:slew}
\end{figure}

A zoom-in of Slew 1 is shown in Figure \ref{fig:slew}, where panels a, b and c are zoom-ins of panels b, c  and f in Figure \ref{fig:timeseries}, respectively, and the spacecraft potential ($-V_{\textnormal{ph}}$) is over-plotted on the MIP densities in Figure \ref{fig:slew}b in blue, to be read off the right-hand y-axis.
We note clear decreases of the LAP2 ion slope and drop-outs of the cold electrons on at least a few occasions when the spacecraft briefly leaves the cavity. This could possibly be taken as an indication that LAP2 goes back into the wake when leaving the cavity, suggesting the existence of a wake and implying directed supersonic flow also outside the cavity. The fact that this happens even though the spacecraft attitude does not change could potentially be taken as an indication of a difference in the ion flow direction. However, we note that many other aspects of the environment also change upon leaving the cavity, most notably the spacecraft potential and density (and consequently also the Debye length), which can be expected to impact the extent and dynamics of the wake. 

Looking at a similar slew a few hours later (Slew 2), between about 00:30 and 06:00 on November 21 (cf.\ Figure \ref{fig:timeseries}), when the spacecraft was mostly outside the cavity, we indeed see a more complicated picture. While there was definitely an increased proclivity towards lager ion slopes and electron currents in LAP2 sweeps (Figure \ref{fig:timeseries}b) during this slew, these effects were less stable and consistent than inside the cavity, indicating that LAP2 was only intermittently out of the wake during this slew. Changes in the ion flow direction would have to be both rapid and erratic to account for such volatile wake effects. This could instead be attributed this to a more dynamic and variable wake outside the cavity. Finally, we note the major slew between 11:30 and 16:00 on November 19 (Slew 3), also mostly outside the cavity, when the attitude change was so large as to bring LAP2 to a position w.r.t.\ the flow comparable to that of LAP1 during nominal pointing (and actually prompting some rather prominent wake effects in COPS, as evidenced by the drop-out in observed neutral density). LAP2 sweeps here exhibited the largest ion and electron currents observed during the plotted time interval. The variability was large here too, even though LAP2 should be consistently out of the wake at these attitudes. This likely reflects the high variability of the plasma observed outside the cavity as discussed previously in the context of LAP1 measurements.

\subsection{Statistical survey of all cavity crossings}
\label{sec:res2}

\begin{figure}[t]
\hspace{-0.22\textwidth}
	\includegraphics[width=1.45\textwidth]{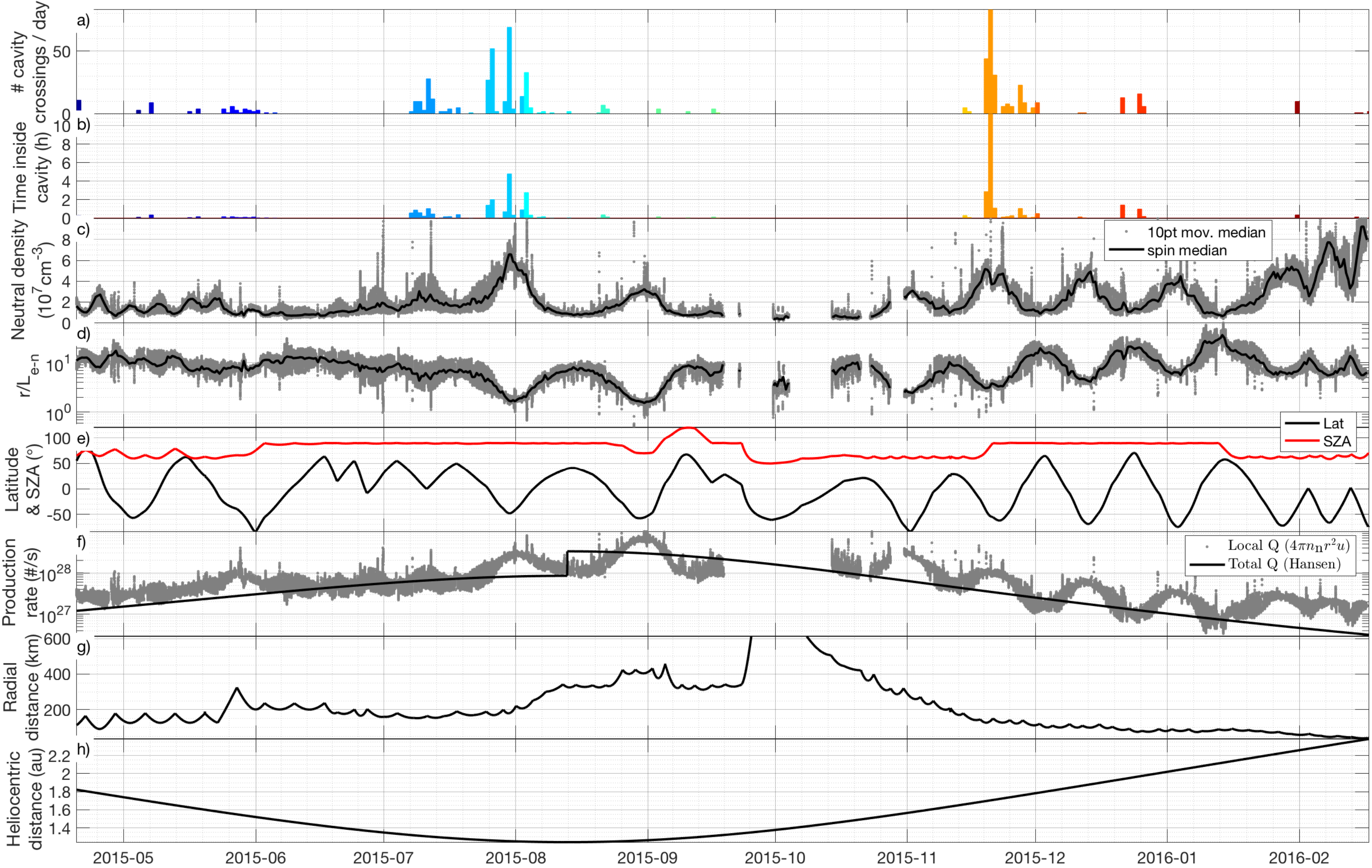}
\caption{Overview of the 300-day period from April 2015 to February 2016 during which all of the cavity crossings occurred. See text for description.}
\label{fig:overview}
\end{figure}

In this Section, we broaden our scope to look at statistics of all cavity crossings throughout the mission. Figure \ref{fig:overview} shows an overview of the 300-day period from April 2015 to February 2016 during which all of the cavity crossings occurred.
Figure \ref{fig:overview}a shows a bar chart of the number of cavity crossings each day during this period (the conspicuous coloring will be explained in connection with Figure \ref{fig:global_vi_hist} below). The distribution is very uneven, with most cavity events occurring in clusters in the end of July, early August, and the end of November 2015, as previously noted by \citet{Goetz2016b} and \citet{Henri2017}. This is even more clear in Figure \ref{fig:overview}b, which shows a bar chart of the total time spent inside the cavity during each day. Here, the Nov 19-21 2015 cluster, examined in detail in the previous Section, really stands out, showing that most of the available data from inside the cavity actually comes from that brief interval. This observational bias is further aggravated by the fact that one of the most prominent days before perihelion, 26 July 2015, is useless for ion velocity measurements since MIP was run in LDL mode throughout most of that day, while the plasma density was likely above the maximum plasma density measurable in LDL mode (about 300 cm-3), so that the actual plasma density was missed by MIP during that period. Figure \ref{fig:overview}c shows neutral density measurements by COPS, filtered by a 10 point moving median filter to remove spurious outliers (grey dots). We have also excluded data collected at spacecraft attitudes corresponding to CAA $>$ 42$^\circ$ or CEA $>$ 90$^\circ$, since COPS appears to be subject to wake-related density drop-outs at these attitudes (cf.\ Section \ref{sec:res1}). In order to bring out the medium to long-term evolution of the neutral density, we also plot the median density for each synodic rotation ($\sim$12 h) of the comet nucleus w.r.t.\ the spacecraft (black line in Figure \ref{fig:overview}c). Figure \ref{fig:overview}d shows the radial distance of the spacecraft in terms of the nominal electron exobase \citep{Henri2017}, denoted $R^*$ and calculated as 

\begin{linenomath*}
\begin{equation}
	R^* = \frac{r}{L_\textnormal{e-n}} = \frac{1}{\sigma_\textnormal{en} n_\textnormal{n} r} \quad , 
\label{eq:exobase}
\end{equation}
\end{linenomath*}
where $\sigma_\textnormal{en}$ is the electron-neutral cross-section, which we have set to $5 \cdot 10^{-16}$ cm$^{-3}$ in accordance with \citet{Itikawa&Mason2005}, $n_\textnormal{n}$ is the neutral density observed by COPS onboard the spacecraft and $r$ is the radial distance to the nucleus. The most prominent clusters of cavity crossings occur at small $R^*$, while many of the minor ones actually at larger $R^*$, as previously noted also by \citet{Henri2017}. We point out here that $R^*$ primarily follows the (reciprocal of the) neutral density, which in turn largely varies in response to changes in the spacecraft latitude, shown in Figure \ref{fig:overview}e (black line), as a consequence of the seasonal variations in outgassing over the comet nucleus. We also note that the conspicuous lack of cavity events at small $R^*$ around the turn of the months Aug-Sep and Oct-Nov can possibly be explained by the low solar zenith angle (SZA, also known as phase angle), shown in Figure \ref{fig:overview}e (red line) at these times; the extent of the cavity is expected to be smaller closer to the Sun-comet line. Figure \ref{fig:overview}f shows the total H$_2$O production rate model by \citet{Hansen2016} (black line) and the local production rate calculated as $4\pi n_{\textnormal{n}}r^2\cdot u$, for a neutral outflow velocity $u$ of 1 km/s (grey dots). (\citeauthor{Hansen2016} actually gives two distinct models for the inbound and outbound passages, hence the discontinuity at perihelion). The most prominent clusters of cavity crossings occur at total production rates between about $3-8 \cdot 10^{27}$~s$^{-1}$. The lack of cavity crossings at higher production rates is likely due to the increased radial distance of the spacecraft during that period. Finally, for context, we also show radial and heliocentric distances in Figures \ref{fig:overview}g and \ref{fig:overview}h, respectively. 

\subsubsection{Ion observations}

\begin{figure}[t]
\center
	\includegraphics[width=0.67\textwidth]{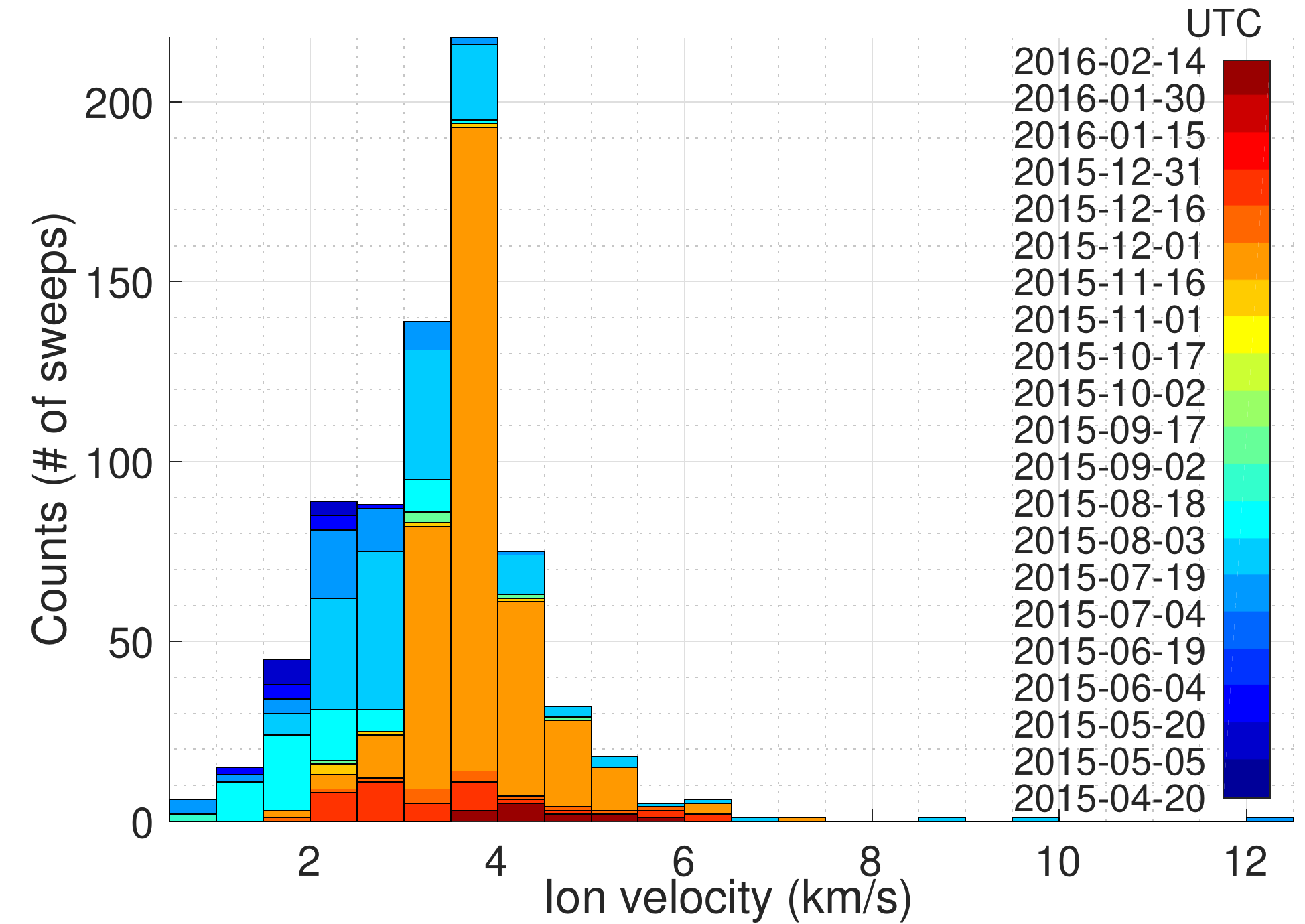}
\caption{Histogram of ion velocity measurements during all diamagnetic cavity crossings throughout the mission, color-coded by time.}
\label{fig:global_vi_hist}
\end{figure}

\begin{figure}[p]
\hspace{-0.22\textwidth}
	\includegraphics[width=1.44\textwidth]{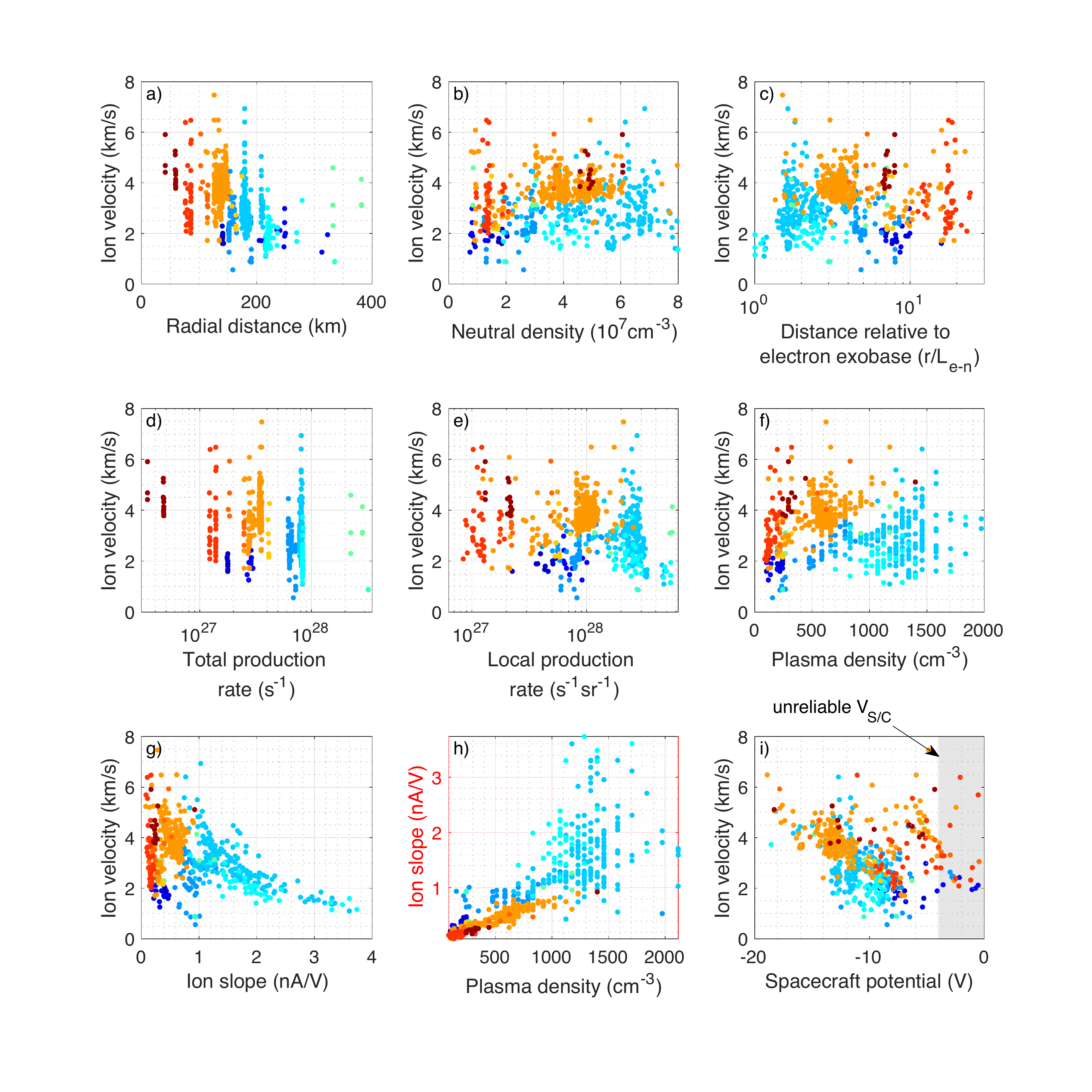}
\caption{Scatter plots of ion velocity vs. a) radial distance b) in situ neutral density c) radial distance relative to electron exobase d) total production rate e) local production rate f) plasma density g) LAP1 ion slopes and i) spacecraft potential, color-coded by time as in Figure \ref{fig:global_vi_hist}. Panel h shows a scatter plot of ion slopes vs. plasma density.}
\label{fig:global_stats}
\end{figure}

Figure \ref{fig:global_vi_hist}a shows a histogram of ion velocity measurements during all diamagnetic cavity crossings throughout the mission, color-coded by time to bring out any long-term temporal variation.
The distribution peaks at velocities around 3.5-4 km/s, although it is clear that this peak derives entirely from the Nov 19-21 2015 cluster. In fact, the rest of the ion velocity measurements distribute more widely between 0-6 km/s, with a central maximum between 2-3 km/s, i.e.\ somewhat lower than the Nov 19-21 cluster. In Figure \ref{fig:global_stats} we examine the data in detail for possible causes of this, by scatter-plotting the ion velocity measurements versus a range of different parameters of interest.
Figure \ref{fig:global_stats}a shows the ion velocity measurements scatter-plotted versus radial distance to the nucleus with the same temporal color-coding as in Figure \ref{fig:global_vi_hist}. This shows an apparent inverse relationship between radial distance and ion velocity. The radial distance generally decreases with time for the cavity crossings as Rosetta could go closer when activity decreased after perihelion, so it is possible that this just reflects an underlying dependence on some other temporally varying parameter. The most prominent deviation from monotonically decreasing distance is the interval 3-18 Aug 2015, when the radial distance was larger than the preceding interval 19 July - 3 Aug 2015. The ion velocity estimates from this period (cyan points just above 200 km in Figure \ref{fig:global_stats}a) come out lower than from the preceding interval (light blue points just below 200 km) in accordance with the general inverse radial dependence but in opposition to the general temporal trend, suggesting that the data is indeed better organized by radial distance than time. On the other hand, the red points just below 100 km, which come from the interval 16-31 Dec 2015, instead deviate from the radial trend, so the picture is not clear.

Figures \ref{fig:global_stats}b-e show similar scatter plots of $v_{\textnormal{i}}$ versus in situ neutral density and $R^*$, total (H$_2$O) production rate and local production rate. The in situ neutral density does not organize the ion velocity measurements very well, suggesting that the ion velocity is not determined by a local collisional equilibrium. $R^*$ organizes the data somewhat better; at least the major blue and orange point clouds separate roughly in agreement with a trend towards higher ion velocities at larger $R^*$, qualitatively consistent with the ions being accelerated radially outward by an electric field outside the exobase. However, this is not very convincing in light of the large scatter in Figure \ref{fig:global_stats}c and the fact that velocity observations from the other time intervals are not well organized by this.

The ion velocity does not appear to correlate much with the local production rate, whereas the total production rate organizes data slightly better, although not better than $R^*$ in Figure \ref{fig:global_stats}c. $R^*$ and the local production rate are really just different ways of combining the radial distance and neutral density; the fact that these combinations do not yield any significant improvement in data organization over the radial distance alone suggests that either the radial distance is the main determining factor for the ion velocity or we have not yet found the real underlying cause. Systematic measurement errors are also a possibility to be considered, as will be further discussed in Section \ref{sec:disc}. In Figures \ref{fig:global_stats}f-g, we plot the ion velocity versus the plasma density observed by MIP and LAP1 ion slope, respectively, since these are the underlying parameters from which $v_{\textnormal{i}}$ is obtained. The main blue and orange point clouds separate nicely in Figure \ref{fig:global_stats}f, showing that the plasma density was generally higher during the 19 July - 18 Aug 2015 interval than for the rest of the data set. However, there is a lot of scatter and no clear correlation between $v_{\textnormal{i}}$ and $n_{\textnormal{e}}$ within the different subgroups. Also, the red and dark blue points, obtained at the very beginning and end of the 300-day period of cavity crossings, respectively, do not follow the general trend. We note here that the plasma density distributes similarly to the neutral density in Figure \ref{fig:global_stats}c, a result of the close relationship between $n_{\textnormal{e}}$ and $n_{\textnormal{n}}$ inside the cavity observed by \citet{Henri2017}, though the plasma density seems to have less variance within each subinterval. Thus, the plasma density does not really provide any new information by which to organize the data. The ion slopes in Figure \ref{fig:global_stats}g groups velocity measurements from the different clusters similarly as the plasma density, but organize them better within each subgroup, especially for the 19 July - 18 Aug 2015 (light blue) interval. In Figure \ref{fig:global_stats}h, we plot the ion slope versus the plasma density. The 19 July - 18 Aug 2015 interval stands out here as being well off the general linear trend that organizes the rest of the measurements. Such a linear trend between ion velocity and plasma density would be expected in the case of a constant ion velocity, in which case the slope of this trend can be obtained from the partial derivative of Equation (\ref{eq:ionslope}) w.r.t.\ $n_{\textnormal{i}}$:

\begin{linenomath*}
\begin{equation}
	\frac{\partial^2 I_{\textnormal{i}}}{\partial n_{\textnormal{i}} \partial U_{\textnormal{B}}} = \frac{2\pi a^2 q_{\textnormal{i}}^2}{m_{\textnormal{i}} v_{\textnormal{i}}}.
\label{eq:ionslopeslope}
\end{equation}
\end{linenomath*}
Solving Equation (\ref{eq:ionslopeslope}) for $v_{\textnormal{i}}$ provides an additional way of obtaining the ion velocity:

\begin{linenomath*}
\begin{equation}
	v_{\textnormal{i}} = \frac{2\pi a^2 q_{\textnormal{i}}^2}{m_{\textnormal{i}}} \Biggm/ \frac{\partial^2 I_{\textnormal{i}}}{\partial n_{\textnormal{i}} \partial U_{\textnormal{B}}}.
\label{eq:vi2}
\end{equation}
\end{linenomath*}
The slope of the linear trend in Figure \ref{fig:global_stats}h is roughly $8\cdot 10^{-4}$ nA/Vcm$^{-3}$, giving $v_{\textnormal{i}} \approx 4$ km/s, i.e.\ in good agreement with the point-wise determination for these time intervals. This gives some extra credibility to these measurements, since if the ion slopes were significantly perturbed by the presence of the charged spacecraft, this effect would be expected to vary with spacecraft potential and Debye length and hence with the plasma density, and the linear relationship would likely be distorted. For the 19 July - 18 Aug 2015 interval it appears that there is a substantial amount of scatter towards higher ion slopes that is not matched by similar increases in MIP density as required for a constant ion velocity. However, we cannot say for sure whether this is due to actual variations in ion velocity or measurement inaccuracies.

Finally, Figure \ref{fig:global_stats}i shows a similar scatter plot of ion velocity versus spacecraft potential. The magnitude of the spacecraft potential, which can be observed to roughly follow the plasma density as expected, should be important for the formation and extent of a spacecraft sheath and pre-acceleration of ions before collection by the probe. We note here that spacecraft potential measurements above -4~V are generally unreliable, as will be discussed below. A general trend towards lower ion velocities for lower-magnitude spacecraft potentials can be observed, possibly indicative of a systematic overestimation of the ion velocities due to pre-acceleration of the ions in the potential field of the spacecraft. However, the scatter is large and, as before, the strong dependence of the spacecraft potential on plasma density, which in turn depends on the neutral density, means that effects of any or all of these parameters cannot be disentangled. In addition, $V_{\textnormal{S/C}}$ also depends on the temperature of the warm electrons, which determines the magnitude of the ambipolar electric field and hence the acceleration of the ions, providing another possible explanation of the observed trend in Figure \ref{fig:global_stats}i.

\subsubsection{Spacecraft potential and warm electrons}

\begin{figure}[t!]
\hspace{-0.22\textwidth}
	\includegraphics[width=1.44\textwidth]{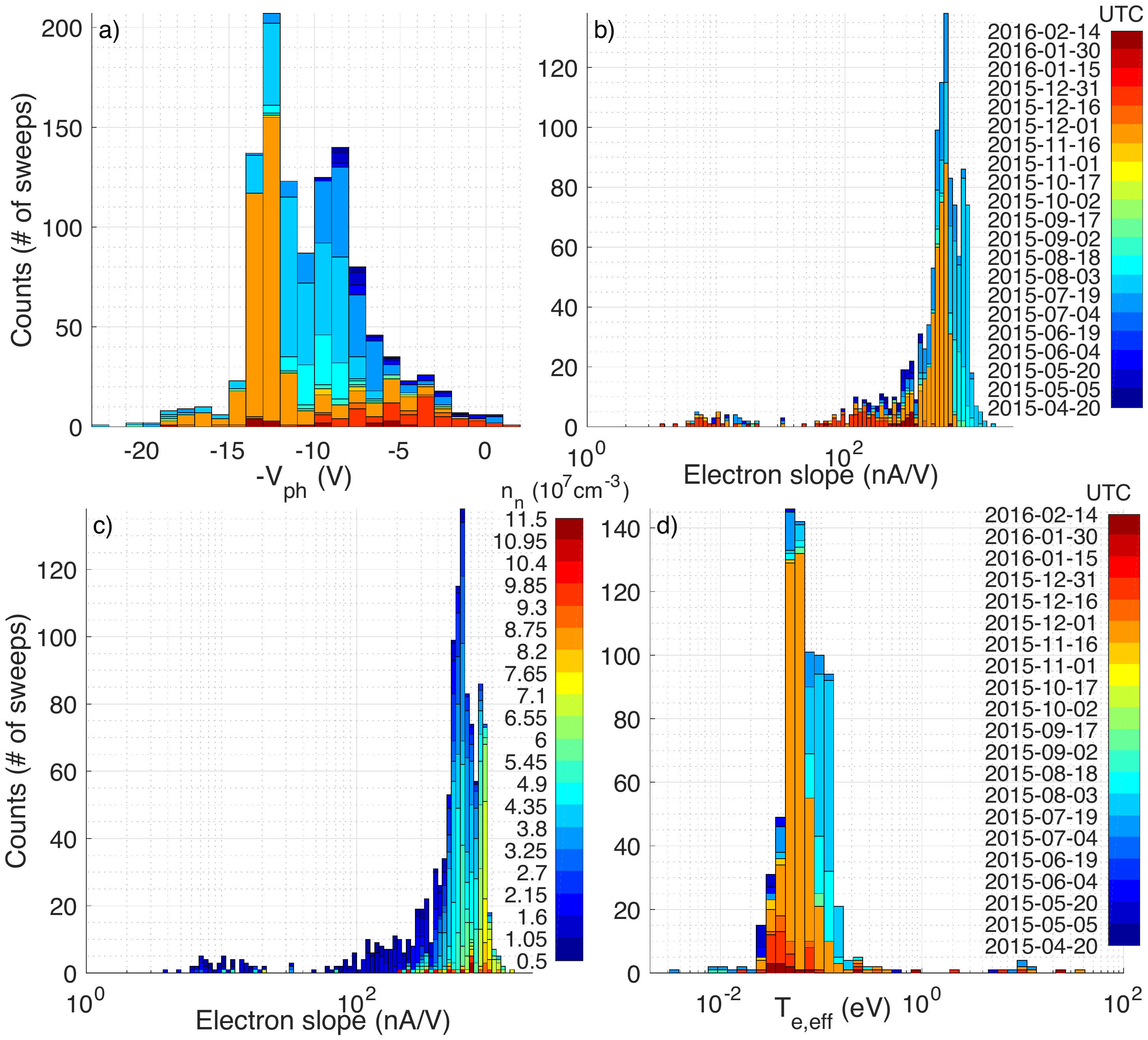}
\caption{Histograms of a) spacecraft potential measurements during all diamagnetic cavity crossings throughout the mission, color-coded by time, b) electron slopes from LAP1 sweeps during all diamagnetic cavity crossings throughout the mission, color-coded by time, c) electron slopes from LAP1 sweeps during all diamagnetic cavity crossings throughout the mission, color-coded by neutral density, and d) effective electron temperatures (see text), color-coded by time, for all LAP1 sweeps obtained during a diamagnetic cavity crossing and for which concurrent MIP density estimates are available.}
\label{fig:global_hists}
\end{figure}

Figure \ref{fig:global_hists}a shows a histogram of spacecraft potential estimates from all 1138 LAP1 sweeps obtained inside the diamagnetic cavity throughout the mission, color-coded by observation time as before. Virtually all of the spacecraft potential values were negative, by at least a few volts. The automated algorithm used to find the location of the photoemission knee in the sweeps is not always very precise, errors of several volts are not uncommon (cf.\ Appendix A). In previous studies \citep{Odelstad2015,Odelstad2017} this was not a major concern due to the long-term, statistical nature of these studies. The more limited and detailed study of the diamagnetic cavity in this paper however calls for a more detailed analysis. Consequently, all sweeps with $V_{\textnormal{ph}} \leq 4$~V in Figure \ref{fig:global_hists}a have been manually double-checked, to find that in all these cases, $V_{\textnormal{ph}}$ was indeed imprecisely determined and that none of these sweeps really had $V_{\textnormal{ph}} < 2$~V, and all but a handful conclusively $\gtrsim$4~V. In addition, $V_{\textnormal{ph}}$ is expected to underestimate the magnitude of $V_{\textnormal{S/C}}$ by a factor 0.7-1 \citep{Odelstad2017}, where the lower end of the interval can be expected to hold for low-magnitude $V_{\textnormal{S/C}}$, as those just discussed, which presumably correspond to low densities and consequently long Debye lengths and inefficient screening of the spacecraft potential. Thus, we conclude that $V_{\textnormal{S/C}}$ was persistently $\lesssim -5$~V throughout all Rosetta's passages through the cavity, attesting to the persistent presence of a warm ($\sim$5~eV) population of electrons all through the parts of the cavity probed by Rosetta.

\subsubsection{Cold electrons}

Figure \ref{fig:global_hists}b similarly shows a histogram of electron slopes from all LAP1 sweeps inside the cavity. Sweeps with clear signatures of cold electrons (electron slope $\gtrsim 100$~nA/V) dominate almost entirely inside the cavity. However, unlike the Nov 19-21 interval showed in Section \ref{sec:res1}, there is a non-negligible number (49) of sweeps lacking the steep-slope signature of cold electrons. These originate predominantly, but not exclusively, from the Dec 2015 time period. Non-detection by LAP does not preclude the presence of cold electrons, since their reaching the probe hinges on the probe bias potential being positive enough to overcome the potential barrier caused by the negatively charged spacecraft (cf.\ Appendix A). Indeed, manual analysis reveals that a handful of the sweeps lacking signatures of cold electrons occur at unusually strong spacecraft potentials, where this could possibly be the case. Also, if the steep-slope part commences very close to the upper edge of the sweep, the electron slope as derived here may not capture it (cf.\ Appendix A); a few cases have been found manually where this was the case. However, some 40 sweeps remain inside the cavity where a cold electron signature is conspicuously absent. About three quarters of these sweeps have concurrent MIP spectra; a manual analysis of these spectra has revealed one case with a clear cold-electron signature and about ten cases where cold the cold-electron signature was clearly absent. The remaining cases were inconclusive due to instrumental effects. Figure \ref{fig:global_hists}c shows the electron slope histogram color-coded by neutral density. All the sweeps lacking a cold-electron signature occur at very low neutral densities (in fact all but one of the points with cyan to green coloring and electron slopes $\lesssim 20$ nA/V in Figure \ref{fig:global_hists}c occur at unusually strong spacecraft potentials or have steep-slope parts that commence too close to the sweep edge to be captured by the electron slope estimate used here, as described above, and the last one can probably be attributed to a spurious COPS measurement), where inefficient cooling might be expected. 

Using Equation (\ref{eq:eslope2}), effective electron temperatures can be calculated from the electron slopes for all LAP1 sweeps inside the diamagnetic cavity for which simultaneous MIP density measurements are available. The electron slopes being a combination of the slopes due to the cold and warm electron populations, this gives only an effective temperature representing their combined contribution to the probe current, although this will be heavily weighted towards the temperature of the cold population, even if its relative abundance is not large. Thus, the effective temperature $T_{\textnormal{e,eff}}$ should be a reasonably accurate estimate of the temperature of the cold electrons, such that $T_{\textnormal{e,cold}} \lesssim T_{\textnormal{e,eff}}$. Figure \ref{fig:global_hists}d shows a histogram of the resulting effective electron temperatures, color-coded by time as in many of the previous figures. They distribute almost entirely in the range 0.02 - 0.2 eV ($\sim 200 - 2000$~K) with a peak between 0.04 - 0.06 eV ($\sim 500 - 700$~K), expected for cometary electrons collisionally cooled by the neutral gas. Both LAP electron slopes and MIP density measurements enter squared in Equation (\ref{eq:eslope2}), thus $T_{\textnormal{e,eff}}$ can be expected to be rather sensitive to errors in either of these quantities. Measurements outside the 0.02 - 0.2 eV range should likely not be trusted. Values $\gtrsim 1$ eV additionally often derive from sweeps without the steep slopes associated with cold electrons, in which case the resulting effective electron temperatures should rather be associated with the warm population. However, these slopes are likely to be affected by potential barrier and spacecraft sheath effects and should not be trusted, though we note that the cluster of values around 5 - 10 eV in Figure \ref{fig:global_hists}d does fall squarely in the expected range for this population. The steep slopes associated with the cold electron population should be less sensitive to potential barrier and spacecraft sheath effects, as suggested by \citet{Olson2010}. From the color-coding in Figure \ref{fig:global_hists}d, a clear temporal trend can be observed, with $T_{\textnormal{e,eff}}$ generally decreasing over time. An effort has been made to determine the cause of this, but the results are as yet inconclusive and will not be further discussed here. This issue should be addressed in more detail in future works.

\section{Discussion}
\label{sec:disc}

\subsection{Uncertainties}

\subsubsection{MIP uncertainties}

In this study, the plasma frequency is identified as the frequency of peak power in MIP spectra. While this is consistent with mutual impedance probe models in plasma characterized by small enough Debye lengths (compared to the transmitter-receiver distance), the plasma frequency is expected to be located below the mutual impedance spectral power peak for larger Debye length (closer to the transmitter-receiver distance) \citep{Geiswiller2001,Gilet2017}. The MIP density estimates may therefore be slightly overestimated. The magnitude of this over-estimation varies from spectrum to spectrum, depending on the Debye length, being virtually negligible for high densities (small Debye lengths) and up to several tens of percent for low densities (larger Debye length), depending on the actual operational mode used for each measurement. A detailed investigation into the interpretation of MIP spectra is beyond the scope of this paper; we note simply that by Equation (\ref{eq:ionslope}), observed ion velocities will be over-estimated if MIP densities are, with the same relative errors ($v_{\textnormal{i}}$ being directly proportional to $n_{\textnormal{i}}$ in Equation (\ref{eq:ionslope})). However, such errors on the order of several tens of percent do not affect the general statistics and conclusions of this paper.


\subsubsection{LAP uncertainties}

LAP1 ion currents are subject to distortions due to the proximity of the large negatively charged spacecraft. First of all, the bias potential $U_{\textnormal{B}}$ w.r.t.\ the ambient plasma is shifted by the spacecraft potential, the probe being grounded to the spacecraft. This effect does not affect the ion (or electron) slopes, since the slope of a linear curve is invariant under translation. (Its effect on other sweep parameters can also largely be compensated for if the spacecraft potential is known, e.g.\ from $-V_{\textnormal{ph}}$). Secondly, ion trajectories are likely strongly perturbed in the potential field of the spacecraft. The spacecraft potential ($\lesssim$-5 V) is generally greater in magnitude than typical ion energies ($\lesssim$ 2 eV) so effects can be expected also upstream of the spacecraft, where LAP1 is located, even though the ion flow is supersonic. Investigations of this are ongoing, but preliminary results from PIC simulations using the SPIS (Spacecraft Plasma Interaction System, \citet{Spis-Sci}) software package indicate that $v_{\textnormal{i}}$ obtained from Equation (\ref{eq:ionslope}) is overestimated as a result of these effects. The magnitude of the overestimation depends on the spacecraft potential, possibly becoming as large as a factor of 2.5 for spacecraft potentials around $-20$ V ($n_{\textnormal{e}} \sim 10^3$ cm$^{-3}$). However, this study is ongoing and possibly overestimates this effect due to uncertainties regarding the ability of the simulations to accurately reproduce the dynamics of flowing low-energy ions, so this should be taken as an upper limit of this error. 
We note that even an error of this magnitude is insufficient to bring our ion velocity estimates down to presumed neutral velocities $\lesssim$1~km/s.

%
%
%

\subsubsection{Ion mass uncertainty}

The ion velocities in this paper have been derived from Equation (\ref{eq:ionslope}) with the assumption of an ion mass $m_{\textnormal{i}} = 18$~u, typical of H$_2$O$^+$ and H$_3$O$^+$ ions presumed to dominate the ion composition. The main other candidate species is CO$_2^+$ with a mass of 44 u, i.e.\ about a factor 2.4 higher. An underestimation of the ion mass in Equation (\ref{eq:ionslope}) would lead to an overestimation of the ion velocity derived from observed ion slopes by the same factor. In the presence of multiple species, the harmonic mean of the individual masses, weighted by their relative abundances, is the relevant quantity for ion velocity determinations by Equation (\ref{eq:ionslope}). The CO$_2$/H$_2$O abundance ratio of the neutral gas did not exceed 1 until after the northward equinox in March 2016 \citep{Gasc2017,Fougere2016b}, i.e.\ after the time period studied here. This does not necessarily constrain the relative abundance of CO$_2^+$ in the plasma since the photoionization frequency of CO$_2$ is about twice that of H$_2$O \citep{Vigren&Galand2013}. On the other hand, CO$_2^+$ has quite a large cross section for electron charge exchange with water \citep{Vigren&Galand2013} which should help suppress the CO$_2^+$ abundance. We have adapted the analytical model of \citet{Vigren2018} to estimate the fractional abundance of CO$_2^+$ ions as a function of the total outgassing rate, an assumed CO$_2$/H$_2$O ratio and an assumed expansion velocity. The relative abundance of CO$_2^+$ at Rosetta's position increases with CO$_2$/H$_2$O ratio, outgassing velocity, and the spacecraft radial distance from the nucleus, but decreases with the outgassing rate. For the two intervals contributing most of the measurements in this study, 19 July - 3 Aug 2015 and 19-21 Nov 2015, CO$_2$/H$_2$O  ratios were $\lesssim 0.1$ and $\lesssim 0.5$, respectively \citep{Fougere2016b}. Global production rates from \citet{Hansen2016} (cf.\ Figure \ref{fig:overview}) were $\gtrsim 1\cdot 10^{27}$~s$^{-1}$. Assuming outgassing velocities (of the neutrals) of $\lesssim 1$~km/s gives (harmonic) mean ion masses of $\lesssim 22$~u and $\lesssim 20$~u, respectively, corresponding to overestimations of $v_{\textnormal{i}}$ by $\lesssim 22$\% and $\lesssim 11$\%, respectively. This is too small to affect the overall statistics, although we cannot rule out that there could be single events at lower outgassing rate and larger CO$_2$/H$_2$O ratio where errors could be larger, though errors in excess of 50\% are hard to achieve by any reasonable parameters using this model, even for single events.

\subsection{Ion drift speed from a simple flux conservation model}

\begin{figure}[t]
\center
	\includegraphics[width=0.67\textwidth]{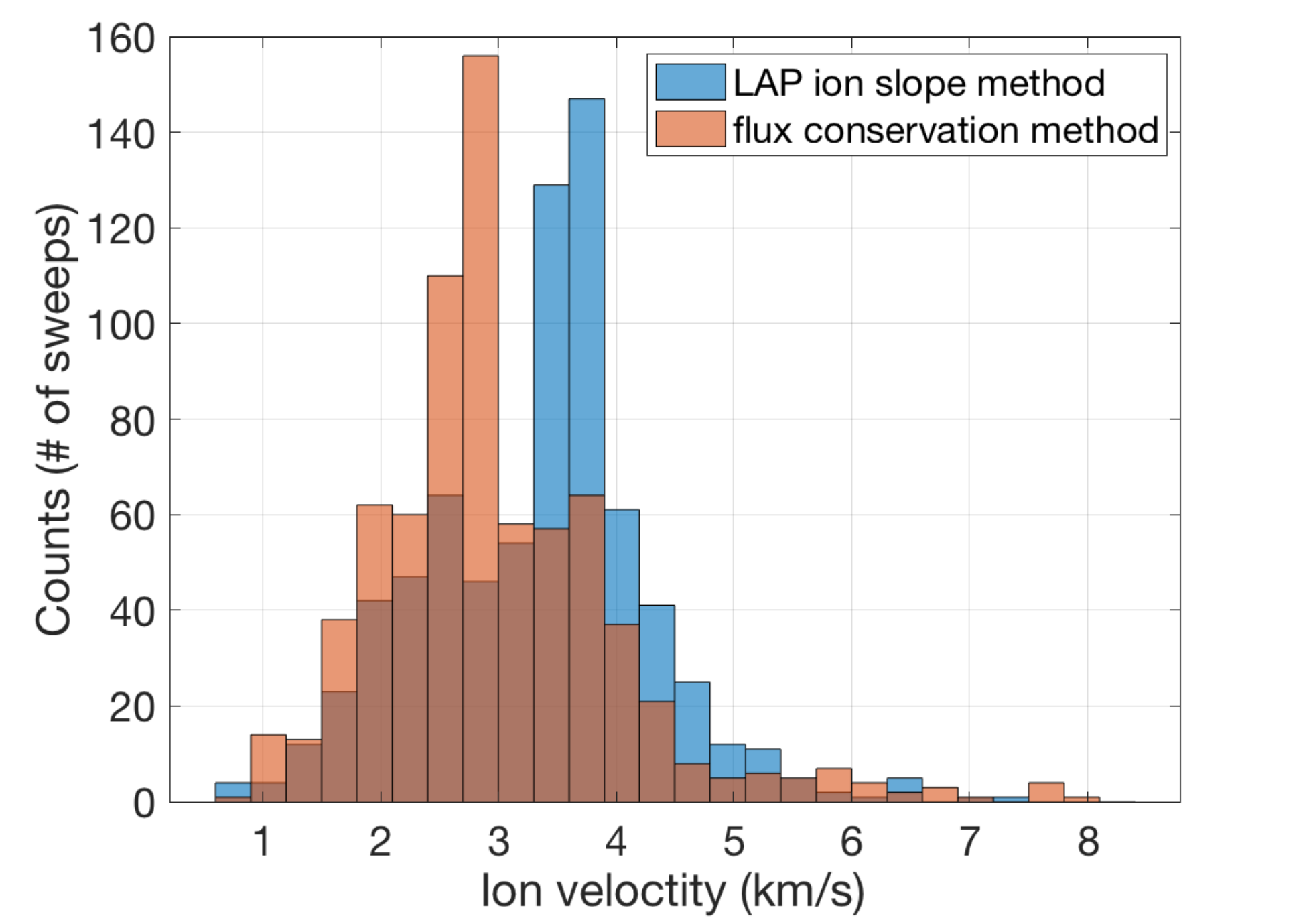}
\caption{Comparison of effective ion drift speeds obtained from the flux conservation model of \citet{Vigren2017b} and the LAP ion slope method during all diamagnetic cavity crossings throughout the mission.}
\label{fig:global_vi_ui_hists}
\end{figure}

\citet{Vigren2017b} proposed a method for obtaining an estimate of the effective ion drift speed from a simple flux conservation model assuming radial outflow. In steady state, the total production of plasma inside the position of the spacecraft must then equal the radial flux of plasma past the spacecraft. Assuming production through photoionization only and neglecting recombination loss, flux conservation gives

\begin{equation}
	n_{\textnormal{i}} u_{\textnormal{i}} = n_{\textnormal{n}} \nu (r - r_{\textnormal{c}}) \quad \Rightarrow \quad u_{\textnormal{i}} = \frac{n_{\textnormal{n}} \nu (r - r_{\textnormal{c}})}{n_{\textnormal{i}}},
\label{eq:flux_conservation}
\end{equation}
where $\nu$ is the photoionization frequency and $r_{\textnormal{c}}$ is the nucleus radius, which is small compared to the spacecraft radial distance throughout the time period covered in this study and can therefore be neglected. Assuming quasi-neutrality, $n_{\textnormal{i}} \approx n_{\textnormal{e}}$, Equation (\ref{eq:flux_conservation}) can be used to obtain estimates of the effective ion drift speed from COPS neutral density and MIP electron density measurements. This provides a different ion velocity estimate, completely independent of LAP measurements and any errors therein, that captures specifically the radial outflow velocity, without any thermal or non-radial contributions. The assumptions of steady state, non-radial outflow and neglect of electron-impact ionization by supra-thermal electrons are well-founded inside the diamagnetic cavity, while the neglect of recombination loss is less clear, leading to a possible overestimation of the ion velocity on the order of a few tens of percent \citep{Vigren2017b,Vigren2015}. Outside the cavity, where the plasma is much more variable and dynamic and the ion motion is affected by the presence of a magnetic field, this velocity estimate is more dubious. We use it here for comparison to and validation of the ion velocities previously obtained inside the diamagnetic cavity by combining MIP densities with LAP ion slopes. We use daily averaged photo-ionization frequencies for H$_2$O, computed at the location of 67P from the Timed SEE L3 v12 database, corrected for phase shift and heliocentric distance, from \citet{Heritier2018}. Figure \ref{fig:global_vi_ui_hists} shows a histogram of the resulting effective ion drift speeds inside the cavity, together with the $v_{\textnormal{i}}$ values derived from LAP ion slopes.
The effective ion drift speeds from the flux conservation model are in good agreement with the ion-slope derived values throughout the cavity. Some shift of the peak of the distribution towards lower velocities can be observed, but our main result that the ion drift velocity is significantly elevated w.r.t.\ the neutral velocity is supported by these measurements.

\subsection{Implications for the pressure balance at the cavity boundary}

%

The fact that we observe ion velocities well in excess of the neutral velocity both inside the cavity and in the surrounding region shows that the ions were not collisionally coupled to the neutrals, implying that the ion-neutral drag force was not responsible for balancing the outside magnetic pressure at the cavity boundary. The poor collisional coupling is a result of the lower outgassing rate of 67P compared to comet 1P/Halley \citep{Vigren2017}, where ion-neutral drag was indeed found to prevail \citep{Ip1987}. In this context it is interesting that \citet{Timar2017} managed to achieve such good fit between observed cavity distance values and the ion-neutral drag model of \citet{Cravens1986,Cravens1987}, which assumes that the ions are collisionally coupled to the neutrals and stagnant at (or just outside) the cavity boundary. The good fit may essentially express the strong dependence of observed cavity boundary distances on the neutral production rate and the varying solar wind pressure. It cannot be ruled out that some other process, driven by the same inputs, might be able to produce an equally good fit, if the resulting dependence of the cavity boundary distance on outgassing and solar wind pressure is similar to that of the ion-neutral drag model.

The elevated ion velocities observed here produce an ion dynamic pressure $n_{\textnormal{i}} m_{\textnormal{i}} v_{\textnormal{i}}^2$ of $\sim$5$\cdot 10^{-10}$ kgm$^{-1}$s$^{-2}$ ($n_{\textnormal{i}} = 1000$ cm$^{-3}$, $m_{\textnormal{i}} = 18$ u\ and $v_{\textnormal{i}} = 4$ km/s). Typical magnetic field strengths outside the cavity are $\sim$20 nT, giving a magnetic pressure $B^2/(2\mu_0) \sim 2 \cdot 10^{-10}$ kgm$^{-1}$s$^{-2}$. Thus the ion dynamic pressure is of the right order of magnitude to balance the magnetic pressure on the outside. If this was the case, the ion dynamic pressure would be expected to decrease outside the cavity, by an amount comparable to the magnitude of the magnetic pressure. We see no such decrease in ion velocity outside the cavity, rather there is some tendency toward increased velocity. While the ion velocity measurements presented in this paper are incapable of distinguishing between bulk and thermal motion and don't give a flow direction, the apparent persistence of a spacecraft wake outside the cavity suggests that no significant thermalization of the ions take place. Since the geometry of the slews during which LAP2 comes out of the wake appear to be generally consistent with radial flow from the nucleus, even outside the cavity, there does not seem to be any clear change in flow direction of the ions either. However, the plasma density generally increases immediately outside the cavity \citep{Henri2017,Harja2018} thus the flux conservation model of \citet{Vigren2017b}, if applied there, would give a decrease in the effective radial ion drift speed and of the ion dynamic pressure. This observation is hard to reconcile with the ion slope based velocity measurements without invoking thermalization or non-radial flow of the ions, which in turn is hard to reconcile with the apparent persistence of a spacecraft wake outside the cavity. The dynamics of the plasma on the outside of the cavity boundary and the role of the ion dynamic pressure in balancing the magnetic pressure should be further investigated in future works.

We have seen that there is always a substantial population of warm ($\sim$5 eV) electrons present throughout the parts of the inner coma probed by Rosetta, so also inside the diamagnetic cavity. These contribute a thermal pressure $n_{\textnormal{e}} k_{\textnormal{B}} T_{\textnormal{e}} \sim 8 \cdot 10^{-10}$ kgm$^{-1}$s$^{-2}$ ($n_{\textnormal{e}} = 1000$ cm$^{-3}$, $T_{\textnormal{e}} = 5$ eV), i.e.\ of the same order of magnitude as the ion dynamic pressure and well sufficient to maintain the pressure balance at the cavity boundary. However, the fact that the electron density increases immediately outside the cavity means that a substantial decrease in electron temperature would be required there. It is hard to conceive of a mechanism that can achieve this in the region immediately outside the cavity boundary. Direct measurements of the electron temperature have not yet been obtained from the data at sufficient accuracy to detect such changes, so the role of the electron thermal pressure for the pressure balance at the cavity boundary cannot be conclusively determined at this time.


\section{Summary and conclusions}
\label{sec:conclusions}

We have combined density measurements by MIP with ion slopes from LAP sweeps to produce measurements of the ion velocity in and around the diamagnetic cavity of 67P. As noted before by \citet{Henri2017}, the density inside the cavity was remarkably stable, while the surrounding region was characterized by high density variations. This turned out to hold also for the ion slopes; they were much more variable outside than inside the cavity. The resulting ion velocity measurements inherited this feature, with stable velocities narrowly distributed around 3.5-4 km/s inside the cavity and scattered toward higher velocities ($\lesssim$8-10 km/s, though still heavily weighted towards 4 km/s) in the surrounding region. The ion velocity inside the cavity was clearly elevated w.r.t.\ the neutral velocity of $\lesssim$1 km/s and in good agreement with the model predictions of \citet{Vigren2017}. This indicates that the ions were not collisionally coupled to the neutrals and implies that the ion-neutral drag force was not responsible for balancing the outside magnetic pressure at the cavity boundary. It also suggests the existence of an ambipolar electric field to accelerate the ions, at least inside the cavity.


Clear evidence of wake effects on LAP2 has been identified inside the cavity during a spacecraft slew. The geometry of this slew was such that LAP2 was brought forward from a position close to being obscured from view of the nucleus to a position more exposed to radial flow from the comet. Inside the cavity, the effect of this on LAP2 was to increase its collection of ions and cold electrons and generally make its current-voltage characteristics more similar to that of LAP1, which was consistently well-exposed to any radial flows. This shows that the flow of cometary ions inside the cavity was indeed close to radial and supersonic, at least w.r.t.\ the perpendicular temperature. Thus, the observed ion velocities inside the cavity are to be taken as a radial velocity of the ions, possibly a cold radial drift, although a non-negligible radial temperature of the ions cannot be rule out. 
Outside the cavity, the effects of these slews on the LAP2 current were less clear and consistent, indicative of a more dynamic wake, or possibly a more variable ion flow direction.

We also made a detailed examination of the spacecraft potential measurements inside the cavity, finding it to be consistently $\lesssim$-5 V. This proves that a population of warm ($\sim$5-10 eV) electrons was present throughout the parts of the cavity probed by Rosetta. This was shown to hold on larger time scales throughout the mission already by \citet{Odelstad2017}, but has now been confirmed on a sweep-by-sweep level for all cavity events. This shows that Rosetta never entered the region of collisionally coupled electrons presumed to exist in the innermost part of the coma, not even during any of the passes through the diamagnetic cavity.

Finally, we have found that a population of cold ($\lesssim$ 0.1 eV) electrons, first shown by \citet{Eriksson2017} to be intermittently present at Rosetta, is in fact observed consistently throughout the diamagnetic cavity. Already immediately outside the cavity, sweeps lacking a signature of cold electrons begin to turn up intermittently. This reinforces the notion of \citet{Henri2017} that the formation and extent of the cavity is closely related to electron-neutral collisionality and suggests that the the region of collisionally coupled electrons, while never entered by Rosetta, was possibly not that far away during the cavity crossings. It also suggests that the filamentation of the cold electrons begins at or near the cavity boundary and is possibly related to an instability of this boundary.

\appendix
\section{Sweep analysis}
\label{sec:sweep_an}

\begin{figure}[t]
	\includegraphics[width=\textwidth]{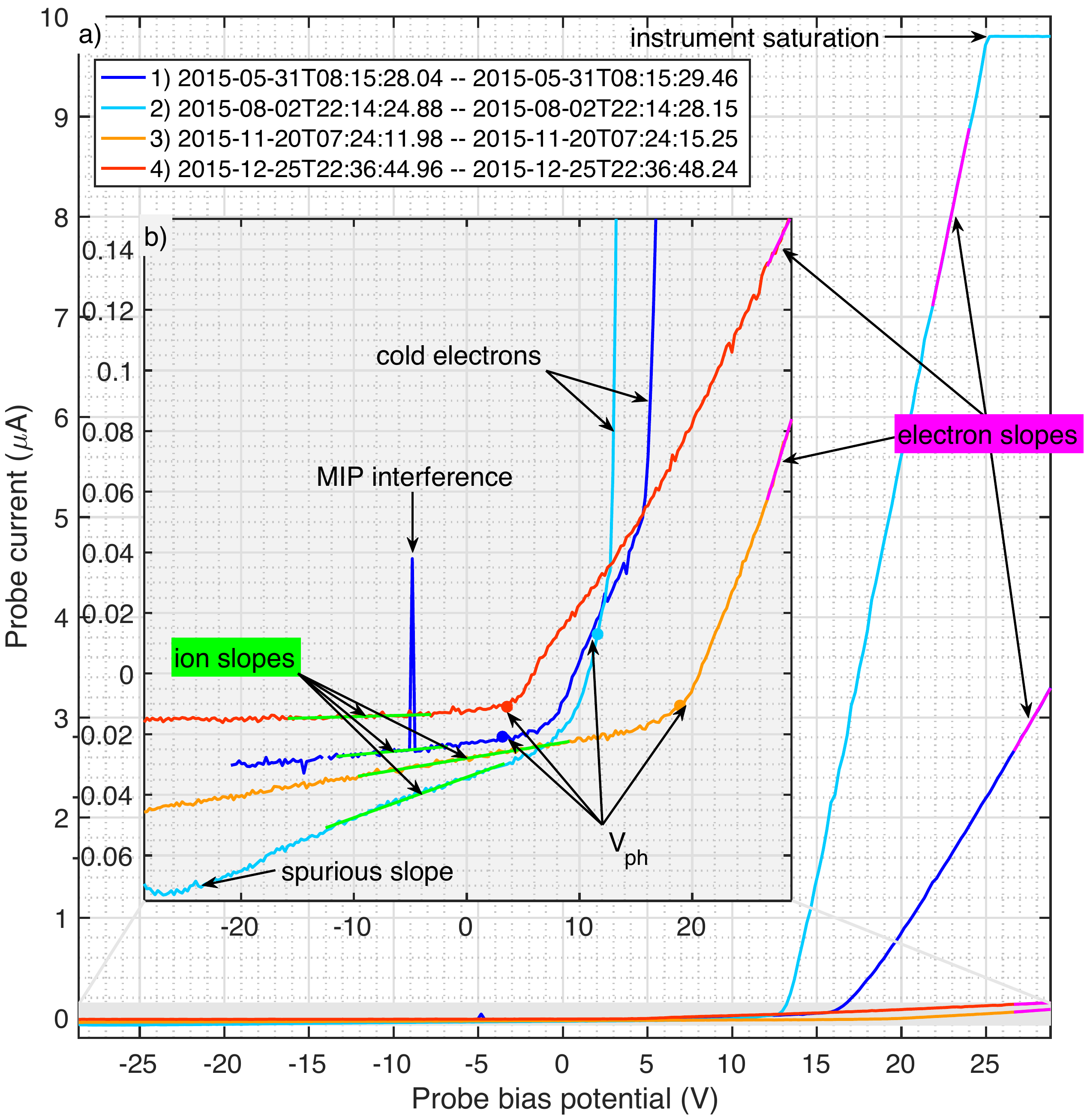}
	\caption{Four examples of LAP1 sweeps from inside the diamagnetic cavity, illustrating the effect of cold electrons and how the electron and ion slopes and the photoelectron knee ($V_{\textnormal{ph}}$) are obtained.}
	\label{fig:ex_sweeps}
\end{figure}

Figure \ref{fig:ex_sweeps} shows four examples of LAP1 sweeps from inside the diamagnetic cavity. 
Together, these four sweeps contain examples of all the major sweep features of relevance to the present analysis and of which account needs to be taken in deriving relevant parameters (e.g.\ electron and ion slopes) from the sweeps. Due to there often being a large difference in scales between various sweep features, the complete sweeps, plotted in the outer axes (Figure \ref{fig:ex_sweeps}a), are complemented by an inset panel (Figure \ref{fig:ex_sweeps}b) displaying zoomed versions of the low-current parts of the sweeps. Here, sweeps 3 and 4 exhibit, at least qualitatively, the behavior expected for spherical probes in a single-temperature maxwellian plasma: The sweeps are linear at negative bias potentials, corresponding to attracted-ion current only, have a sharp inflection point at the photoelectron knee potential ($V_{\textnormal{ph}}$), corresponding to the local plasma potential w.r.t.\ the spacecraft (which may differ from the plasma potential at infinity due to some fraction of the spacecraft potential field remaining at the position of the probe), and then become linear again at more positive potentials, corresponding to attracted-electron current only. In sweeps 1 and 2 on the other hand, the linear parts above $V_{\textnormal{ph}}$ are interrupted by secondary inflection points at about 16 and 12 V, respectively, followed by a much steeper slope. This is attributed to a second population of cold ($\lesssim$0.1 eV) electrons \citep{Eriksson2017}, required to produce such steep slopes in the sweeps. The fact that they don't appear in the sweeps until at least a few volts above $V_{\textnormal{ph}}$ we ascribe to the existence of a potential barrier between the probe and the ambient plasma due to some fraction of the potential field of the negatively charge spacecraft remaining outside the position of the probe \citep{Olson2010}. The probe would then have to go to somewhat more positive potentials than the local plasma potential in its immediate vicinity in the potential well of the spacecraft for this barrier potential to be fully suppressed, allowing the cold electrons to reach the probe. Assuming this barrier potential to be much lower in magnitude than typical temperatures of the uncooled thermal electron population, which must exist to explain the persistently negative spacecraft potentials \citep{Odelstad2017}, the bulk of the thermal electrons will have sufficient energy to overcome it and will appear in the sweeps already at the photoelectron knee potential.

The photoelectron knee is identified in the sweeps primarily by seeking peaks in the sweep second derivative. This requires considerable filtering and pre-processing of the sweeps to eliminate noise and spurious effects, without thereby smoothing the knee so much that it avoids subsequent detection. Further analysis includes conditioning on various supplementary sweep parameters, such as the sign of the current at candidate knees, to single out real from spurious knees and distinguish between the photoelectron knee and the secondary knee associated with collection of cold electrons. The algorithm currently in use is complex and will not be further described here. For the purpose of this study, it suffices to note that the resulting $V_{\textnormal{ph}}$ estimates have typical errors up a few volts (as can be observed for sweeps 1 and 2 in Figure \ref{fig:ex_sweeps}b).

The electron slopes are obtained from an ordinary least squares (OLS) linear fit to the last 10 points of each sweep that have currents below 9~$\mu$A. This cap is used since in some operational modes, the LAP electronics saturate at currents above this value, as for example sweep 4 in Figure \ref{fig:ex_sweeps}a. The limited range of only 10 points for the fit is used to assure that in the presence of cold electrons the fit is performed well above the second inflection point, so that the two very different slopes below and above this point are not convoluted. This generally works well, but for some rare cases where the second inflection point appears within 10 points of the sweep upper edge and the electron slope is consequently underestimated. As the linear electron part is the steepest on the sweeps, we can never overestimate the slope (except for small random errors). The resulting linear fits are shown in cyan for the sweeps in Figure \ref{fig:ex_sweeps}.

The ion slopes are obtained from similar linear fits to the part of each sweep lying between 40\% and 80\% of the distance up from the first (most negative) bias potential to the photoelectron knee ($V_{\textnormal{ph}}$). In previous work where sweep ion slopes have been used, either directly \citep{Odelstad2017,Vigren2017b}, or indirectly, they have been obtained from the first 40\% of each sweep below $V_{\textnormal{ph}}$, i.e.\ just below the part used here. This choice was made to ensure that the repelled electron current had fully decayed and did not skew the ion slopes. In this study, we have found numerous examples of spurious features in these parts of the sweeps, one example of which is shown in sweep 4 in Figure \ref{fig:ex_sweeps}b, where there is a clear bend in the sweeps at the most negative bias potentials. Therefore, the ion slopes in this paper are derived from further up the sweep, where such effects have not been observed and which are more consistently linear. We surmise that the electron current is well suppressed in this interval and that the previous selection was unnecessarily draconian in this regard.

In LDL mode, interference from MIP produces spurious outlying points in the part of the sweeps used to obtain the ion slope. An example of this can be seen in sweep 3 in Figure \ref{fig:ex_sweeps}b. To remedy this, a generalized extreme Studentized deviate (ESD) test \citep{Rosner1983} is iteratively performed on sequential fits, testing for and removing one outlier at a time, until no outliers (with respect to a normal distribution and at a significance level of 0.001) are found in the fit residuals. (This approach to outlier removal was used together with the total linear least squares (TLS) algorithm by \citet{Odelstad2017}; we use it here together with the OLS algorithm instead and with a factor 10 lower significance level.) This procedure has been found to generally produce accurate and robust linear fits to the aforementioned parts of the sweeps, even in the presence of such strong interferences from MIP. It is used blindly on all sweeps in this paper; no effort has been made to selectively apply it only to operational modes where MIP interferences can be expected (LDL modes, cf.\ Section \ref{sec:MIP}). The resulting linear fits are shown in green for the sweeps in Figure \ref{fig:ex_sweeps}.

\acknowledgments
Rosetta is a European Space Agency (ESA) mission with contributions from its member states and the National Aeronautics and Space Administration (NASA). The work on RPC-LAP data was funded by the Swedish National Space Board under contracts 109/12, 168/15 and 166/14 and Vetenskapsr{\aa}det under contract 621-2013-4191. Work at LPC2E was supported by CNES, by ESEP and by ANR under the financial agreement ANR-15-CE31-0009-01. This work has made use of the AMDA and RPC Quicklook database, provided by a collaboration between the Centre de Donn{\'e}es de la Physique des Plasmas (CDPP) (supported by CNRS, CNES, Observatoire de Paris and Universit{\'e} Paul Sabatier, Toulouse) and Imperial College London (supported by the UK Science and Technology Facilities Council). Work at the University of Bern on ROSINA COPS was funded by the State of Bern, the Swiss National Science Foundation, and the European Space Agency PRODEX Program. The data used in this paper will soon be made available on the ESA Planetary Science Archive and is available upon request until that time.






%
%
%
%
%
%
%
%
%
%

\bibliographystyle{agufull08}
\bibliography{references}{}





\listofchanges

\end{document}